\documentclass[apj]{emulateapj}

\usepackage{amsmath}
\usepackage[varg]{txfonts} 

\usepackage{graphicx}
\usepackage[outdir=.]{epstopdf} 

\usepackage[german,british]{babel}
\usepackage{biblio}
\usepackage{bibentry}
\usepackage{natbib}

\usepackage{color}
\usepackage{microtype}

\usepackage[]{hyperref}

\hypersetup{%
  colorlinks=true,
  linkcolor=red,
  citecolor=blue,
  urlcolor=blue,
  bookmarksopen=true,
  bookmarksnumbered=true,
  pdfstartpage={1},  
  pdftitle = {},
  pdfsubject = {},
  pdfauthor = {} ,
  pdfkeywords = {},
  pdfproducer = {LaTeX with hyperref}
}

\usepackage{etoolbox}

\makeatletter

\patchcmd{\NAT@citex}
 {\@citea\NAT@hyper@{%
    \NAT@nmfmt{\NAT@nm}%
    \hyper@natlinkbreak{\NAT@aysep\NAT@spacechar}{\@citeb\@extra@b@citeb}%
    \NAT@date}}
 {\@citea\NAT@nmfmt{\NAT@nm}%
  \NAT@aysep\NAT@spacechar\NAT@hyper@{\NAT@date}}{}{}

\patchcmd{\NAT@citex}
 {\@citea\NAT@hyper@{%
    \NAT@nmfmt{\NAT@nm}%
    \hyper@natlinkbreak{\NAT@spacechar\NAT@@open\if*#1*\else#1\NAT@spacechar\fi}%
      {\@citeb\@extra@b@citeb}%
    \NAT@date}}
 {\@citea\NAT@nmfmt{\NAT@nm}%
  \NAT@spacechar\NAT@@open\if*#1*\else#1\NAT@spacechar\fi\NAT@hyper@{\NAT@date}}
 {}{}

\makeatother

\def\nH{\ifmmode{\>n_{\textnormal{\sc h}}} \else{$n_{\textnormal{\sc h}}$}\fi}

\def\erg{\ifmmode{\> {\rm erg}}\else{erg}\fi}
\def\keV{\ifmmode{\> {\rm keV}}\else{keV}\fi}

\def\deg{\ifmmode{\>^{\circ}}\else{$^{\circ}$}\fi}
\def\onedeg{\ifmmode{\>1^{\circ}}\else{$1^{\circ}$}\fi}

\def\xvir{\ifmmode{\>x_{vir}}\else{$x_{vir}$}\fi}
\def\Mvir{\ifmmode{\>M_{vir}}\else{$M_{vir} $}\fi}
\def\rvir{\ifmmode{\>r_{vir}}\else{$r_{vir}$}\fi}
\def\vvir{\ifmmode{\>v_{vir}}\else{$v_{vir}$}\fi}

\def\tratio{\ifmmode{\>\tau}\else{$\tau$}\fi}

\def\rms{\ifmmode{\>r_{\textnormal{\sc ms}}}\else{$r_{\textnormal{\sc ms}}$}\fi}

\def\Mpc{\ifmmode{\>{\rm Mpc}} \else{Mpc}\fi}
\def\kpc{\ifmmode{\>{\rm kpc}} \else{kpc}\fi}
\def\pc{\ifmmode{\>{\rm pc}} \else{pc}\fi}

\def\Gyr{\ifmmode{\>{\rm Gyr}} \else{Gyr}\fi}
\def\Myr{\ifmmode{\>{\rm Myr}} \else{Myr}\fi}
\def\yr{\ifmmode{\>{\rm yr}} \else{yr}\fi}
\def\pyr{\ifmmode{\>{\rm yr}^{-1}}\else{yr $^{-1}$} \fi}
\def\s{\ifmmode{\>{\rm s}}\else{s}\fi}
\def\ps{\ifmmode{\>{\rm s}^{-1}}\else{s$^{-1}$}\fi}
\def\Hz{\ifmmode{\>{\rm Hz}}\else{Hz}\fi}

\def\kms{\ifmmode{\>{\rm km\,s}^{-1}}\else{km~s$^{-1}$}\fi}

\def\K{\ifmmode{\>{\rm K}}\else{K}\fi}

\def\sr{\ifmmode{\>{\rm sr}}\else{sr}\fi}
\def\psr{\ifmmode{\>{\rm sr}^{-1}}\else{sr$^{-1}$}\fi}
\def\arcs{\ifmmode{\>{\rm arcsec}}\else{arcsec}\fi}
\def\parcs{\ifmmode{\>{\rm arcsec}^{-1}}\else{arcsec${-1}$}\fi}
\def\parcss{\ifmmode{\>{\rm arcsec}^{-2}}\else{arcsec${-2}$}\fi}

\def\cm{\ifmmode{\>{\rm cm}}\else{cm}\fi}
\def\cc{\ifmmode{\>{\rm cm}^{3}}\else{cm$^{3}$}\fi}
\def\sqc{\ifmmode{\>{\rm cm}^{2}}\else{cm$^{2}$}\fi}
\def\pcc{\ifmmode{\>{\rm cm}^{-3}}\else{cm$^{-3}$}\fi}
\def\psc{\ifmmode{\>{\rm cm}^{-2}}\else{cm$^{-2}$}\fi}

\def\g{\ifmmode{\>{\rm g}}\else{g}\fi}
\def\Msun{\ifmmode{\>{\rm M}_{\odot}}\else{M$_{\odot}$}\fi}
\def\hMsun{\ifmmode{\> h^{-1}{\rm M}_{\odot}}\else{$h^{-1}$M$_{\odot}$}\fi}

\def\Zsun{\ifmmode{\>{\rm Z}_{\odot}}\else{Z$_{\odot}$}\fi}

\def\rayl{\ifmmode{\>{\rm R}}\else{R}\fi}
\def\mR{\ifmmode{\>{\rm mR}}\else{mR}\fi}

\renewcommand{\ion}[2]{\hbox{#1\,{\sc #2}}}
\newcommand{\los}{sightline}

\def\lya{\ifmmode{\>{\rm Ly}\alpha}\else{Ly$\alpha$}\fi}

\def\Ha{\ifmmode{\>{\rm H}\alpha}\else{H$\alpha$}\fi}
\def\Hb{\ifmmode{\>{\rm H}\beta}\else{H$\beta$}\fi}

\def\HI{\ifmmode{\> \textnormal{\ion{H}{i}}} \else{\ion{H}{i}}\fi}
\def\HII{\ifmmode{\> \textnormal{\ion{H}{ii}}} \else{\ion{H}{ii}}\fi}
\def\CIV{\ifmmode{\> \textnormal{\ion{C}{iv}}} \else{\ion{C}{iv}}\fi}
\def\SiIV{\ifmmode{\> \textnormal{\ion{S}{iv}}} \else{\ion{Si}{iv}}\fi}

\def\NHI{\ifmmode{\> {\rm N}_{\HI}} \else{N$_{\HI}$}\fi}
\def\MHI{\ifmmode{\> {\rm M}_{ \HI}} \else{M$_{\HI}$}\fi}

\def\mua{\ifmmode{\>\mu_{ \textnormal{\Ha}}}\else{$\mu_{ \textnormal{\Ha}}$}\fi}
\def\alphabha{\ifmmode{\>\alpha_{B}^{(\textnormal{\Ha})}}\else{$\alpha_{B}^{(\textnormal{\Ha})}$}\fi}

\newcommand{\fyra}{\textsc{fyris alpha}}

\slugcomment{Revised version 1}

\shorttitle{Accretion onto the Galactic halo}
\shortauthors{T.~Tepper-Garc\'\i{}a, J.~Bland-Hawthorn, and R.~S.~Sutherland}

\defcitealias{pla14b}{\textcolor{black}{Planck Collaboration} 2014}

\begin{document}


\title{The Magellanic Stream: break up and accretion onto the hot Galactic corona}


\author{Thor Tepper-Garc\'\i{}a\altaffilmark{1}, Joss Bland-Hawthorn\altaffilmark{1}, and Ralph S.~Sutherland\altaffilmark{2}}
\affil{$^1$Sydney Institute for Astronomy, School of Physics, University of Sydney, NSW 2006, Australia}
\affil{$^2$Mount Stromlo Observatory, Australia National University, Woden, ACT 2611, Australia}

\begin{abstract}

The Magellanic \HI\ Stream ($\approx 2\times 10^9 \Msun\; [d/55\kpc]^2$) encircling the Galaxy at a distance $d$ is arguably the most important tracer of what happens to gas accreting onto a disk galaxy. Recent observations reveal that the Stream's mass is in fact dominated (3:1) by its ionised component. Here we revisit the origin of the mysterious \Ha\ recombination emission observed along much of its length that is overly bright ($\sim 150-200\mR$) for the known Galactic ultraviolet (UV) background ($\approx 20-40\mR\; [d/55\kpc]^{-2}$).
In an earlier model, we proposed that a slow shock cascade was operating along the Stream due to its interaction with the extended Galactic hot corona. We find that, for a smooth coronal density profile, this model can explain the bright \Ha\ emission if the coronal  density satisfies $2 \times 10^{-4} < (n / \pcc) < 4 \times 10^{-4}$ at $d = 55$ kpc. But in view of updated parameters for the Galactic halo and mounting evidence that most of the Stream must lie far beyond the Magellanic Clouds ($d>55\kpc$), we revisit the shock cascade model in detail. At lower densities, the \HI\ gas is broken down by the shock cascade but mostly mixes with the hot corona without significant recombination. At higher densities, the hot coronal mass (including the other baryonic components) exceeds the baryon budget of the Galaxy. If the \Ha\ emission arises from the shock cascade, the upper limit on the smooth coronal density constrains the Stream's mean distance to $\lesssim 75$ kpc. If, as some models indicate, the Stream is even further out, either the shock cascade is operating in a regime where the corona is substantially mass-loaded with recent gas debris, or an entirely different ionization mechanism is responsible.
\end{abstract}


\section{Introduction} \label{sec:intro}

	The Galaxy is surrounded by a vast amount of neutral gas in the form of high-velocity \HI\ clouds \citep[HVC;][]{oor70a}. Formally, these are neutral gas structures at a Galactic latitude $\vert b \vert > 30^{\circ}$ having kinematic properties not consistent with the overall Galactic rotation \citep[][]{wak01a}. We now recognize that many of these make up the Magellanic Stream \citep[MS;][]{die71a,wan72a,mat74a}, roughly $2\times 10^9 \Msun\; [d/55\kpc]^2$ of gas \citep{fox14a} that has been stripped from the Magellanic Clouds (MCs), two dwarf galaxies in orbit around the Galaxy at a mean distance $d \approx 55$ kpc \citep{wal12a,gra14a}. 

	The Magellanic Stream is ideal to study the environment of the Galaxy. Radio (\HI\ 21cm) surveys show that the Stream extends for 200$^\circ$ across the Southern Galactic Hemisphere \citep{nid10a}, and absorption line measurements towards distant quasars indicate a cross-section of roughly one quarter of the whole sky \citep{fox14a}. Dynamical models \citep[][]{bes07a,gug14a} constrained by accurate measurements of the proper motions of the MCs \citep{kal13a} agree that they move on a highly eccentric orbit, and that the MS spans a wide range in Galactocentric distance, from its source in the MCs system at roughly $55\kpc$ to $ 80 - 150 \kpc$ above the South Galactic Pole (SGP) all the way to the tip of the tail.

Given its relative proximity, the MS has been observed across the electromagnetic spectrum. Beyond \HI, it has been detected in molecular \citep{ric01b} and ionized \citep{lu94a,fox05b,sem03a} gas. A shadowing experiment aimed at measuring the coronal soft X-ray emission discovered that the emission is enhanced in the direction of the MS \citep{bre09a}. Recombination optical emission (\Ha) was detected for the first time by \citet{wei96a} and later confirmed by others \citep[][]{rey98a,put04a,mad12a}. But despite repeated attempts, to date, no stars have been discovered at any location along the Stream \citep[e.g.][]{ost97a}. The data obtained by the recently completed Hubble Space Telescope (HST) Cosmic Origins Spectrograph (COS) UV absorption survey of the MS \citep[][]{fox13a,ric13a,fox14a} indicates that the Stream is dominated by ionized gas, as was first proposed by \cite{bla07a}. These data collectively suggest the existence of a strong interaction between the Stream gas and the hot halo \citep[or corona;][]{spi56a} of the Galaxy.

	Radiative hydrodynamic models \citep[the 'shock cascade';][]{bla07a} indicate that the MS-halo interaction may be strong enough to explain observed disruption of the Stream \citep{nid10a} and its high ionization fraction. At the same time, the presence of coherent and strong enough magnetic fields \citep[][]{put98a} may stabilise the gas against severe ablation and provide thermal insulation to inhibit total evaporation of the neutral clouds \citep{mcc10b}. Whether such a shielding mechanism is operating all along the Stream is currently unknown. But without it, the Stream is likely to evaporate and mass-load the Galactic halo with a substantial amount of baryons \citep{mcc15b}.
	
	Therefore, if we are to understand the complex environment of galaxies and how gas settles into galaxies \citep{hei09b}, we need to explain first the observed properties of the Magellanic Stream within the framework of a multiphase hydrodynamical model. Elucidating the mechanism behind the bright spots of \Ha\ emission observed along the Magellanic Stream has proven to be particularly challenging \citep[see][]{bla07a}. To date, there have been two competing models \citep[see also][]{kon01a}: (i) the slow shock cascade discussed above; (ii) a new interpretation invoking a powerful flare of UV radiation from the Galactic Centre (GC) powered by the accretion of material onto the central black hole in Sgr A$^*$ \citep{bla13a}.
	
	The GC flare model is inspired by two important circumstances: 1) The discovery of the $\gamma$-ray emitting bubbles discovered by the {\em Fermi} satellite extending roughly 50$^\circ$ ($10 \kpc$) from the GC \citep{su10a}; 2) The observation that the brightest optical emission along the Stream is confined to a cone with half-angle $\theta_{1/2} \approx 25^\circ$ roughly centred on the SGP \citep{mad12a}. The GC flare model has found support from the timescales and energy budget required to ionise the Stream, which are consistent with the results from jet-driven numerical models of the {\em Fermi} bubbles \citep{guo12a}. More recently, \cite{fox14a} 
have discovered that the ionisation levels over the SGP require an energetically harder ionising spectrum 
than elsewhere along the Stream, with the exception of a localised region near the LMC.

	The shock cascade model, on the other hand, explains the observation that the brightest \Ha\ detections lie at the leading edges of the \HI\ clouds that make up the MS \citep{wei02a}. However, this model may fail to produce the observed emission levels {\em if} the distance to the Stream at the SGP significantly exceeds 
the traditional view of $d = 55 \kpc$, as indicated by most orbit calculations for the Magellanic Clouds
over the past five years \citep[][]{bes12a,gug14a}. The shock cascade model is strongly dependent on the density structure of the Galactic hot halo, and it assumes that the coronal density smoothly declines with Galactocentric distance as $\propto r^{-2}$. But it now appears that both of
these assumptions may be false justifying our efforts to revisit the shock cascade model.
	
	The goal of this study is to investigate the strength of the recombination (\Ha) emission produced by the interaction of the MS gas with the Galactic corona exploring a range of Galactocentric distances and different halo parameters, i.e., adopting different density profiles and temperatures of the gas sitting at rest in a fixed dark-matter (DM) potential. Note that throughout the paper we assume a flat, dark-energy- and matter (baryonic and cold dark-matter; CDM) dominated Universe, and a cosmology defined by the set of parameters (relevant to this work) $h = 0.7$, $\Omega_m = 0.3$, and $\Omega_{\Lambda} = 0.7$.

\section{A model of the Galactic halo} \label{sec:halo}

The density and temperature structure of the Galactic corona is largely determined by the underlying gravitational potential. The potential, in turn, is determined by the three main components of the Galaxy: the stellar bulge, the stellar and gaseous disc, and the DM halo. Given the mass of the bulge ($\sim 10^{10} \Msun$) and the disc \citep[$\sim10^{11} \Msun$;][]{kaf14a}, and their size, these components are expected to dominate the Galactic potential only at $r \lesssim 5 \kpc$ and at $r \lesssim 15 \kpc$, respectively. In other words, with exception of the inner $\sim 5 \kpc$ \citep{weg15a}, the DM halo dominates the Galactic potential at all distances, with a similar contribution from the disc at scales comparable to its length. The distance to the nearest point of the MS is believed to be
$d \approx 55 \kpc$, which is the average distance to the LMC \citep[$\approx 50 \kpc$;][]{wal12a}, and the SMC \citep[$\approx 60 \kpc$;][]{gra14a}. Consequently, in modelling the interaction of the Magellanic Stream with the Galactic corona, it is safe to ignore the contributions to the potential from the bulge and the disc, and to focus instead on the DM halo only.

\subsection{The Galactic DM halo} \label{sec:dmhalo}

We model the DM halo of the Galaxy assuming it is well described by a single-component isothermal sphere. We opt for such a model given its solid physical foundation, with properties that can be derived from first principles starting from a few basic assumptions \citep[see e.g.][]{bin08a}.

The {\em scale-free} potential, $W$, and the {\em scale-free} density, $y$, of an isotropic isothermal (DM) sphere are defined by \citep[][]{kin66a}:
\begin{align}
	\frac{ 1 }{ x^2 } \frac{ d }{ dx } \left( x^2 \frac{ dW }{ dx}\right) & = -9 ~y_{ \textnormal{\sc iso} }(x) \, , \label{eq:ode1} \\
	y_{ \textnormal{\sc iso} }(x) & = \exp\left[ W(x)\right] \, . \label{eq:ode2} 
\end{align}
where $x$ is a scale-free coordinate.

This system of equations has no analytic solution, but it can be integrated numerically to values $x \ll 1$ to (nearly) arbitrary precision. Appropriate boundary conditions are, for example, the requirement that both the potential and the force vanish at the origin, i.e. $W(0) \equiv 0$ and $W'(0) \equiv 0$, respectively, which implies $y_{ \textnormal{\sc iso} }(0) = 1$.  Note that $W(x) \leq 0$ for $x \geq 0$. 

The connection between the scale-free quantities and their physical counterparts is given by
\begin{equation} \label{eq:iso}
	r = r_c x \, ; \quad \rho(r) = \rho_c y_{ \textnormal{\sc iso} }(x) \, ; \quad \psi(r) = \sigma^2 W(x) \, .
\end{equation}
Here, $\rho_c$, $r_c$, and $\sigma$ are the central density, the core radius, and the constant velocity dispersion, respectively. A solution corresponding to a particular physical system is obtained by fixing two of these three parameters (or any other two independent physical quantities of the system, for that matter); the third parameter is tied to the other two through the relation $r_c^2 = 9 \sigma^2 / 4 \pi G  \rho_c$ \citep{kin66a}.

\begin{figure*}[htbp]
\includegraphics[width=0.48\textwidth]{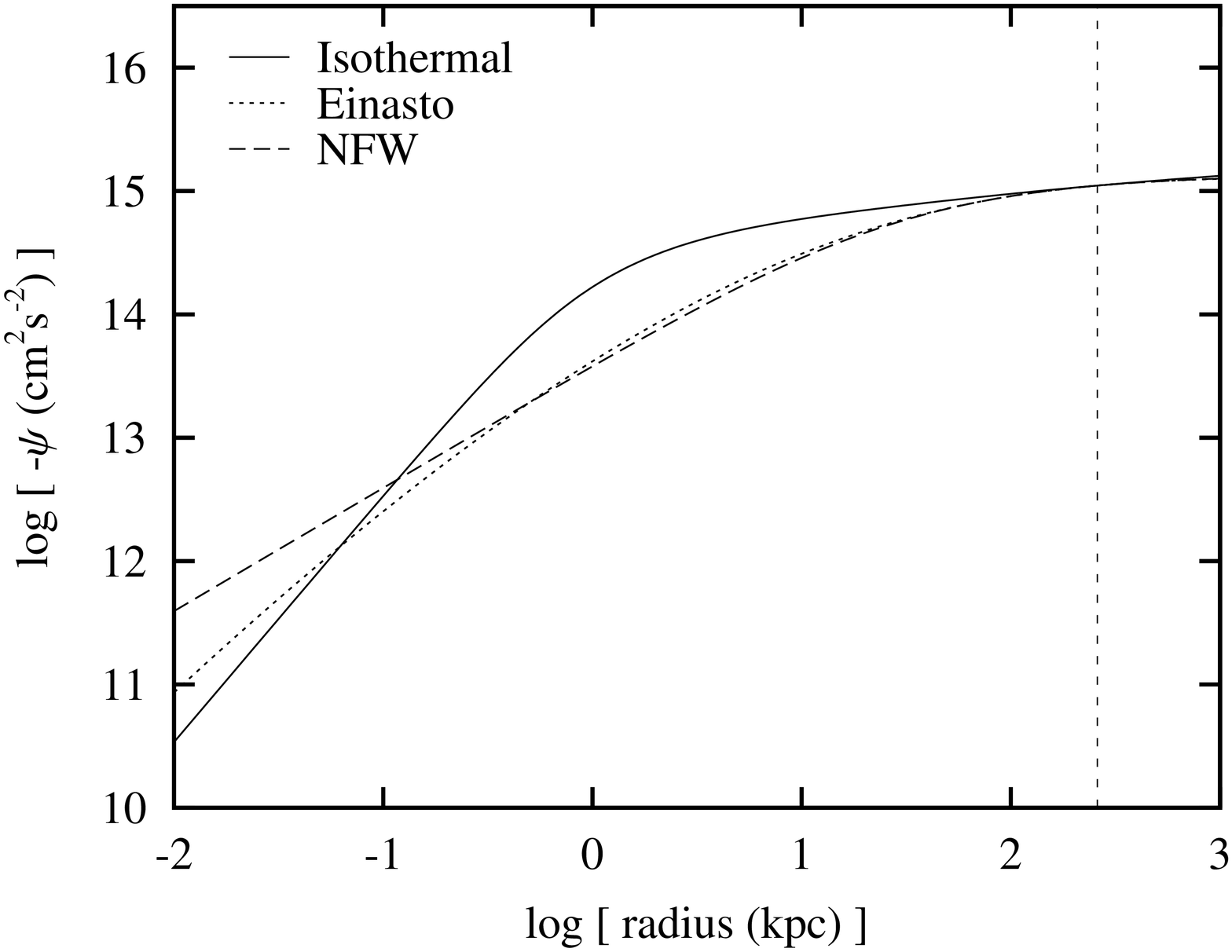}
\hfill
\includegraphics[width=0.48\textwidth]{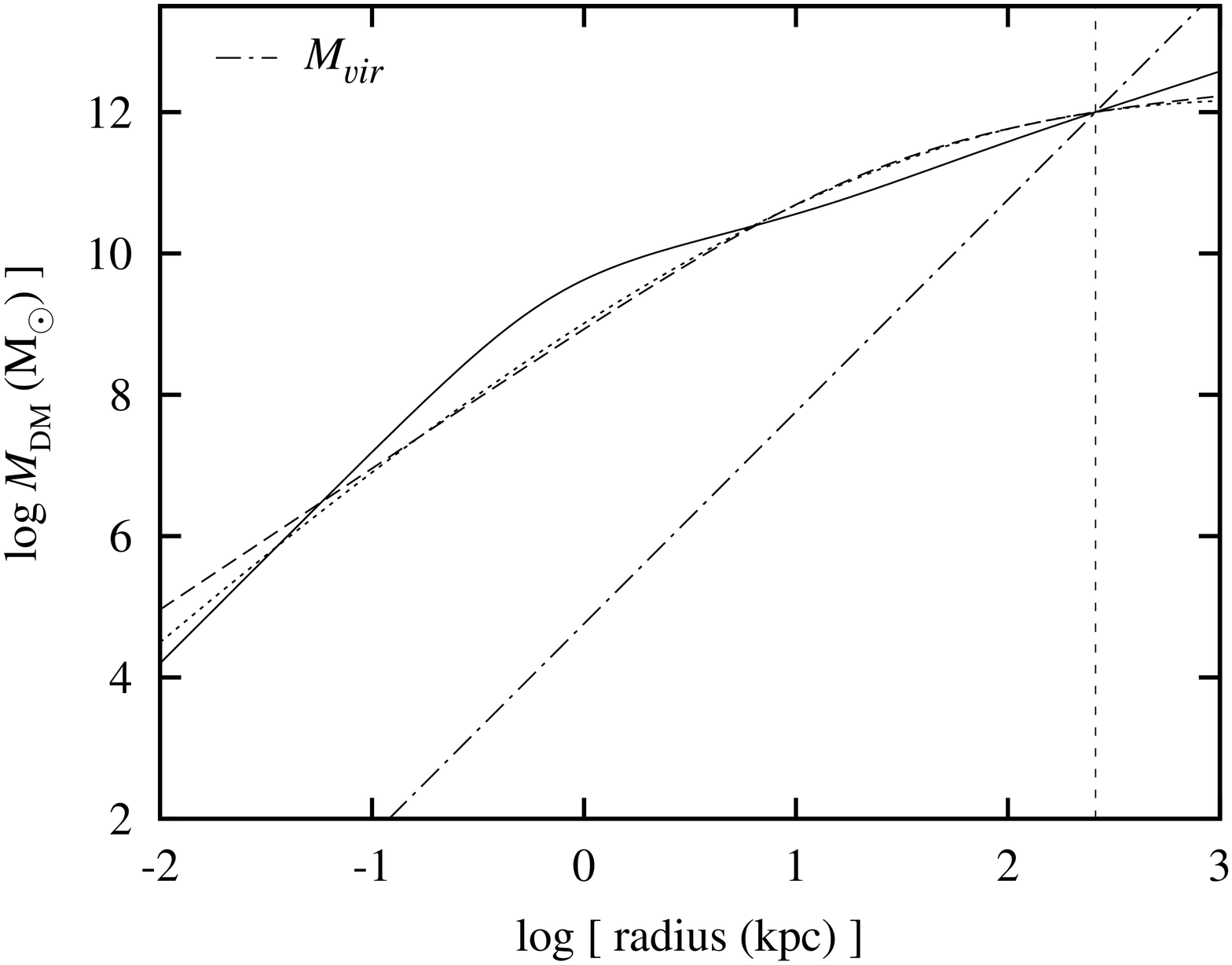}
\caption[ DM potential and DM mass ]{ Potential (left) and enclosed mass (right) of the Galactic DM halo. Each curve corresponds to a different model, indicated by the legend in the left panel. The line of constant slope in the right panel shows the relation $\Mvir(r)$ given by equation \eqref{eq:rvir}. Note that the potential and the mass of all profiles have been matched at the virial radius of the Galaxy, $\rvir \approx 260 \kpc$, indicated by the vertical line in both panels. See also Table \ref{tab:dmhalo}.}
\label{fig:dmhalo}
\end{figure*}

Alternatives to the isothermal sphere as a viable choice to describe a self-gravitating system of collisionless, DM particles, i.e. a DM halo include the \citet*[][NFW]{nav96a} model,
\begin{equation} \label{eq:nfw}
	y_{ \textnormal{\sc nfw } }(r) = (r / r_s)^{-1} (1 + r / r_s)^{-2} \, ,
\end{equation}
where $r_s$ is a characteristic scale length; and the \citet{ein65a} profile,
\begin{equation} \label{eq:ein}
	y_{ \textnormal{\sc ein} }(r) = \exp{ \left\{ - \frac{ 2 }{ \alpha } \left[ \left( \frac{ r }{ r_s } \right)^{ \alpha } - 1 \right] \right\} }
\end{equation}
where $\alpha$ is a free parameter.
Regardless of the model adopted, if the density profile $\rho(r)$ is known, a general formalism can be applied which allows us to calculate scale-free quantities \citep[q.v.][]{ste02b} on a case-by-case basis to describe a particular physical system.

In order to specify our isothermal DM halo, and to compare its properties to the NFW and Einasto models, we proceed as follows. Based on the results of \citet[][]{kaf14a}, who assume the dark halo of the Galaxy to be of NFW type, we fix the virial mass $\Mvir$ and the concentration $\xvir \! \equiv \! \rvir\ \! / r_c$ of the NFW dark halo to $\Mvir \! = 10^{12} \Msun$, and $\xvir = 15$, respectively. Our adopted value for the virial mass implies\footnote{The viral mass and the virial radius are linked to one another through the relation
\begin{equation} \label{eq:rvir}
		\rvir = \left( \frac{3 \Mvir}{ 4 \pi \Delta_c \overline{\rho}_{m} } \right)^{1/3} \, .
\end{equation}
We adopt a value for the cosmic mean matter density $\overline{\rho}_{m} \approx 2.76 \times 10^{-30} \g \pcc$ and $\Delta_c \approx 337$. Note that we define the virial radius at $z = 0$ \citep[cf.][]{shu14a}.
}
$\rvir \approx 260 \kpc$ and $\vvir \equiv  (G \!\Mvir / \!\rvir\ \!)^{1/2} \approx 130  \kms$, respectively.
Then we calculate the scale parameters for the isothermal and the Einasto DM halos by requiring that the virial mass and the {\em physical} potential at the virial radius in each case match the corresponding values for the NFW dark halo. Table \ref{tab:dmhalo} summarises our assumed values (in bold font face) and lists the derived values of the relevant scaling parameters for each of the three halo models. The potential and corresponding mass for each model are shown in Figure \ref{fig:dmhalo}. As can be seen, the relevant properties of an isothermal DM halo are very similar to the NFW and Einasto models. We note that the virial temperature is very similar across models, which is of relevance for the discussion in later sections.

\begin{table}
\begin{center}
\caption{DM Halo Properties}
\label{tab:dmhalo}
\begin{tabular}{lccc}
\hline
\hline
										& Isothermal			& Einasto$^f$	 & NFW			~\\
\hline\\
\Mvir\ [$10^{12}$ \Msun]						& {\bf 1.00}			& {\bf 1.00}	& {\bf 1.00}$^a$	~\\
\rvir\ 	[\kpc]\								& 259				& 259		& 259			~\\
\vvir\	[\kms]\								& 129				& 129		& 129			~\\
\xvir\										& 433				& 13.8		& {\bf 15.0}$^a$	~\\
$\psi_{vir}	~[10^{15} \cm^2 \s^{-2}]$				& {\bf -1.11 }			& {\bf -1.11}	& -1.11			~\\
$r_{s} ~[\!\kpc]$$^b$							& 0.60				& 18.8		& 17.3			~\\
$v_{s} ~[\!\kms]$$^c$						& 90.1				& 82.4		& 104			~\\
$M_{h}(\leq\!\!\rvir)	~[10^{10} \Msun]$$^d$		& 1.41				& 3.03		& 1.32			~\\
$T_{\textnormal{\sc dm}}	~[10^5 \K]$$^e$		& 5.81				& 4.85		& 7.73			~\\
\end{tabular}
\end{center}
\tablecomments{Fixed parameter values are shown in bold face; derived values, in normal font face. See text for details. $^a$Gives the value of the `concentration'. The value \xvir\ = 15 is from \citet[][]{kaf14a}. $^b$Corresponds to the {\em scale} radius for the Einasto and NFW profiles, and to the {\em core} radius $r_c$ for the isothermal sphere. $^c$Corresponds to the {\em scale} velocity for the Einasto and NFW profiles, and to the {\em velocity dispersion} $\sigma$ for the isothermal sphere. $^d$Obtained from equation \eqref{eq:gasm}, assuming $\tratio \equiv 1$ and $\mu = 0.59$. $^e$Obtained from equation \eqref{eq:tdm}, assuming $\mu = 0.59$. $^f$ We adopt $\alpha = 0.18$ \citep[][]{gao08a}.
 }
\end{table}

\begin{figure*}[htbp]
\includegraphics[width=0.48\textwidth]{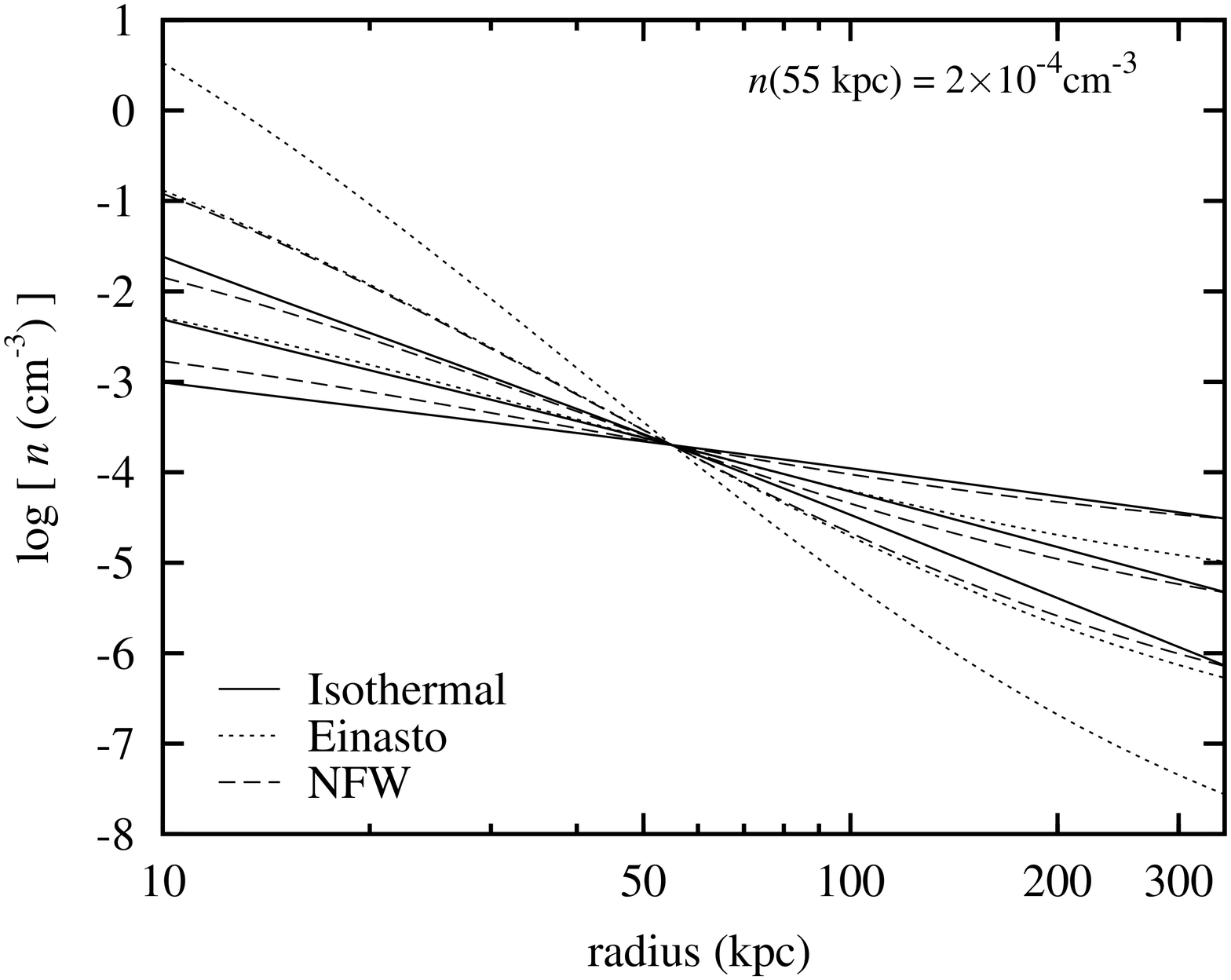}
\hfill
\includegraphics[width=0.48\textwidth]{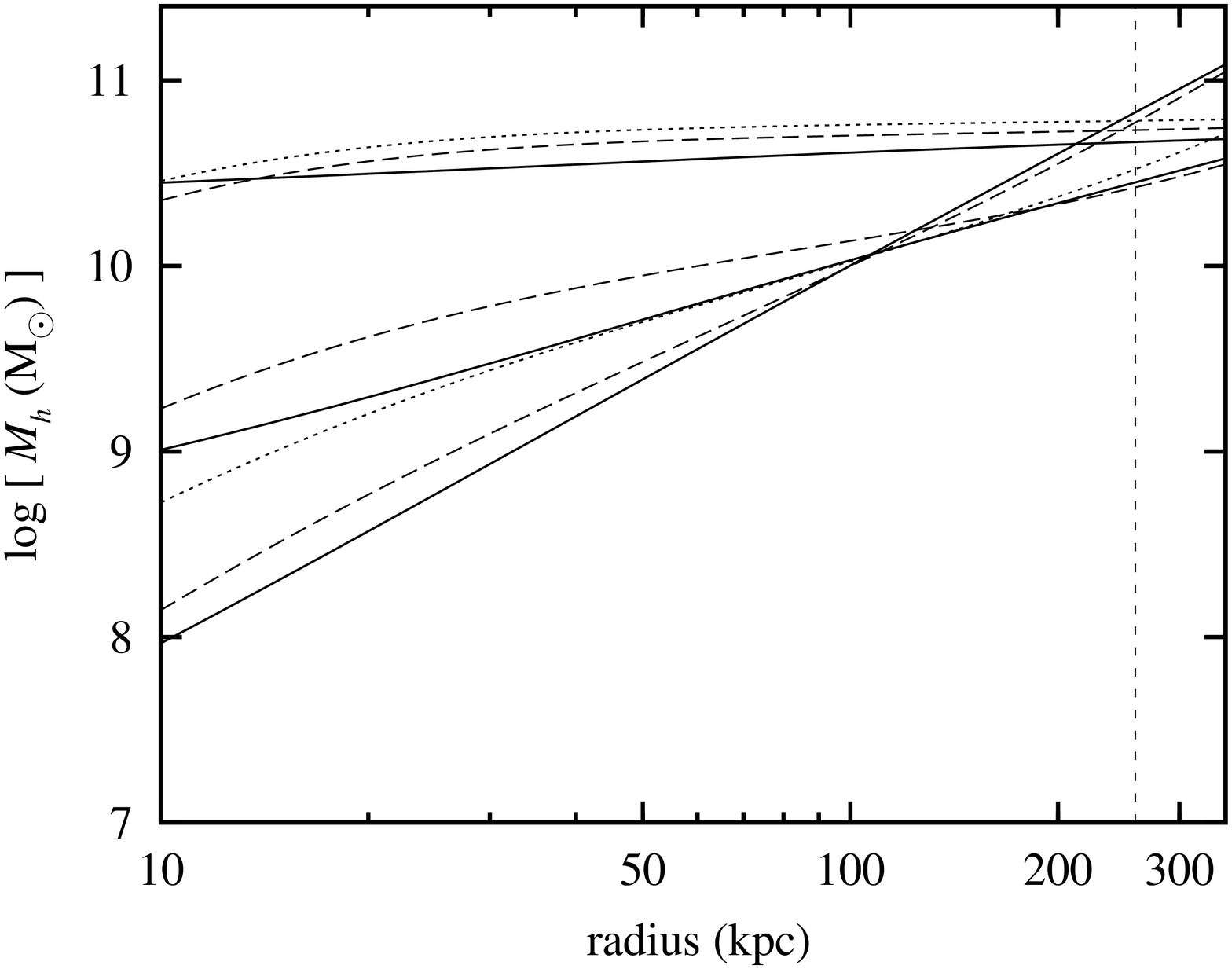}
\caption[  ]{Left: Density profile model (equation \ref{eq:partdens}) adopting different DM halo models, all normalised to $n(55 \kpc) =  2 \times 10^{-4} \pcc$. For a given halo, the curves correspond, from top to bottom, to values of \tratio\ = 1.5, 1.0, and 0.5. Right: Gas mass enclosed within a given radius (equation \ref{eq:gasm}) for each of the density profiles shown in the left panel. Note that the order of the curves is inverted with respect to the order on the left panel, i.e.. for a given halo model, the lowest curve corresponds to  \tratio\ = 1.5, etc. The vertical line indicates the virial radius of the Galaxy, $\rvir \approx 260 \kpc$. Note that the enclosed mass for the Einasto model with \tratio\ = 0.5 is well above $10^{11} \Msun$ and is not shown.}
\label{fig:dmhalo2}
\end{figure*}

\subsection{The Galactic corona} \label{sec:gashalo}

It has long been known that the potential well of the Galaxy is filled with a diffuse, hot, gaseous component \citep[the corona;][]{spi56a}. However, the origin of this gas, its thermodynamic state, its physical properties, its extension and hence its total mass, are still unknown. A recent attempt to constrain the density structure of the corona implies a total gas mass of the Galaxy around $\sim10^{10} \Msun$ within $\rvir$ \citep{mil13a,mil15a}, which is consistent with other estimates \citep[e.g.][]{gat13a}. This important result relies on the assumptions that the corona is smooth, in collisional ionisation equilibrium, and isothermal, with a temperature of $2 \times 10^{6} \K$. While the gaseous halo of the Galaxy is most likely not smooth nor strictly isothermal, the mean temperature of the gas inferred from its associated X-ray emission appear remarkably uniform across the sky \citep{hen13e, hen14e}.

Given this circumstance, we model the hot halo of the Galaxy as a single-phase, {\em smooth}, spherically symmetric component consisting of an ideal gas at a constant temperature $T_{h}$, in hydrostatic equilibrium with the DM potential $\psi$ of an isothermal sphere. We further assume that the self-gravity of the gas is negligible, which is justified in the case of the Galaxy given that the inferred gas mass of the hot halo is on the order of $10^{-2}$ the mass of the DM halo  \citep{sut98a}. Under these assumptions, the total particle density of the corona is given by $n(r) = n_0 \exp[ \psi(r) / a^2 ]$. Here, $a^2 = k T_{h} /  \mu m_u$ is the isothermal sound speed, $\mu$ is the mean molecular weight, $m_u$ is the atomic mass unit, and $k$ is Boltzmann's constant. 

The virial `temperature' of the dark matter halo follows from the equivalence $a^2 \equiv \sigma^2$,
\begin{equation} \label{eq:tdm}
	T_{\textnormal{\sc dm}} \equiv \frac{ \mu \, m_u }{ k } \sigma^2 \, .
\end{equation}
It worth emphasising that the above is is merely an {\em equivalent} temperature, as $\mu \, m_u$ is not literally the DM particle mass (which is currently unknown). Table \ref{tab:dmhalo} lists the virial temperatures of the different models.

It is straightforward to show that under these assumptions the distribution of the hot gas in the potential is effectively governed by the {\em thermal ratio} \citep{cav76a}\footnote{Note that the designation of thermal ratio by the Greek letter $\beta$ is widespread in the literature. Here, we adopt its original designation.}
\begin{equation} \label{eq:tau}
	\tratio \equiv \frac{ T_{\textnormal{\sc dm}} }{ T_{h} } = \sigma^2 / a^2 \, ,
\end{equation}
such that
\begin{equation} \label{eq:partdens}
		n(r) = n_0 \exp{\left[ \!\tratio W(r / r_c)  \right]} \, ,
\end{equation}

The total gas mass within $r$ for a particular value of \tratio\ follows from the integral of $n$ over the appropriate volume, 
\begin{equation} \label{eq:gasm}
		M_{h}(r) =4 \pi  \, \mu ~m_u \int_0^r \!\!\! n(x) x^2 dx \, .
\end{equation}
%

\begin{figure*}[htbp]
\includegraphics[width=0.48\textwidth]{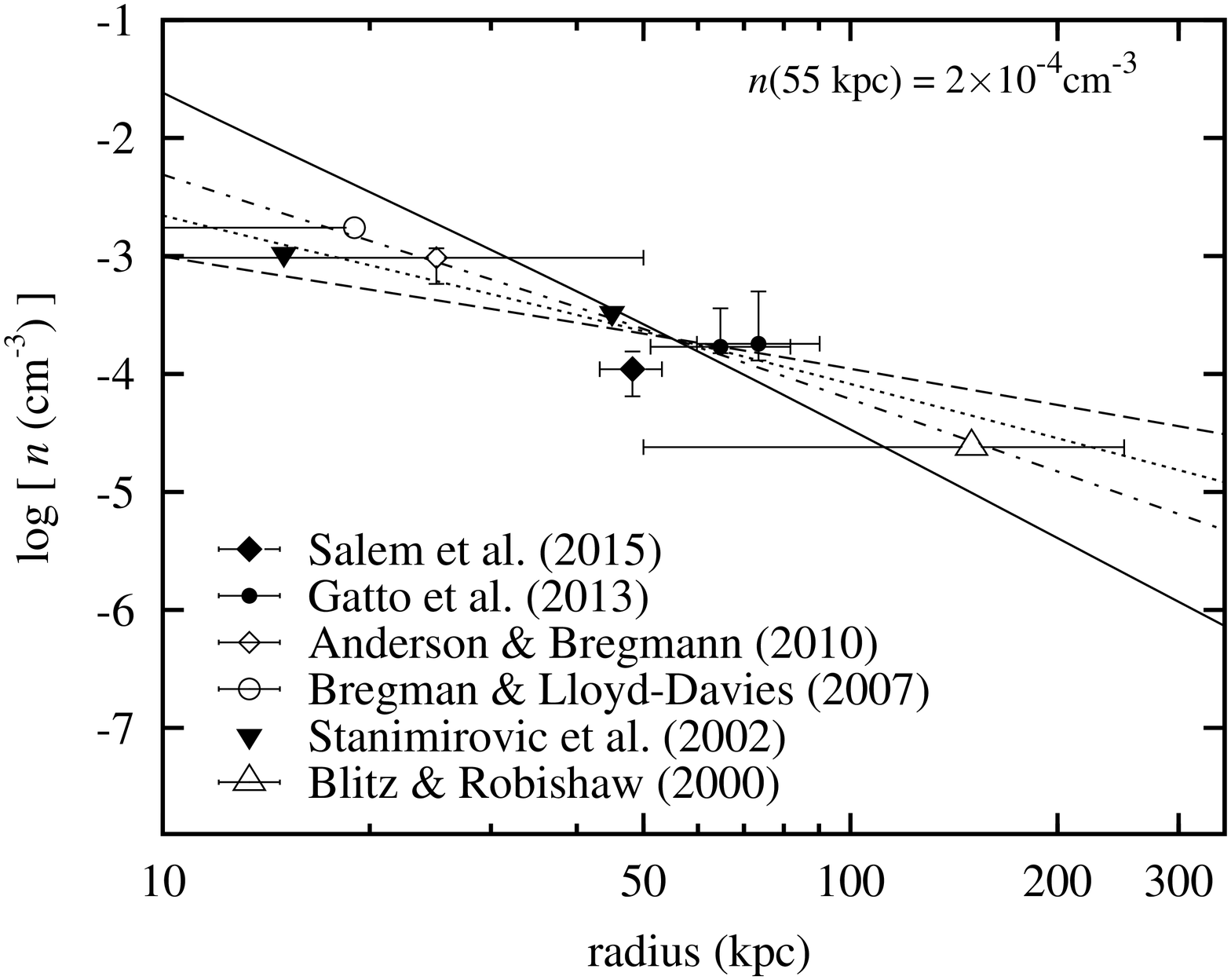}
\hfill
\includegraphics[width=0.48\textwidth]{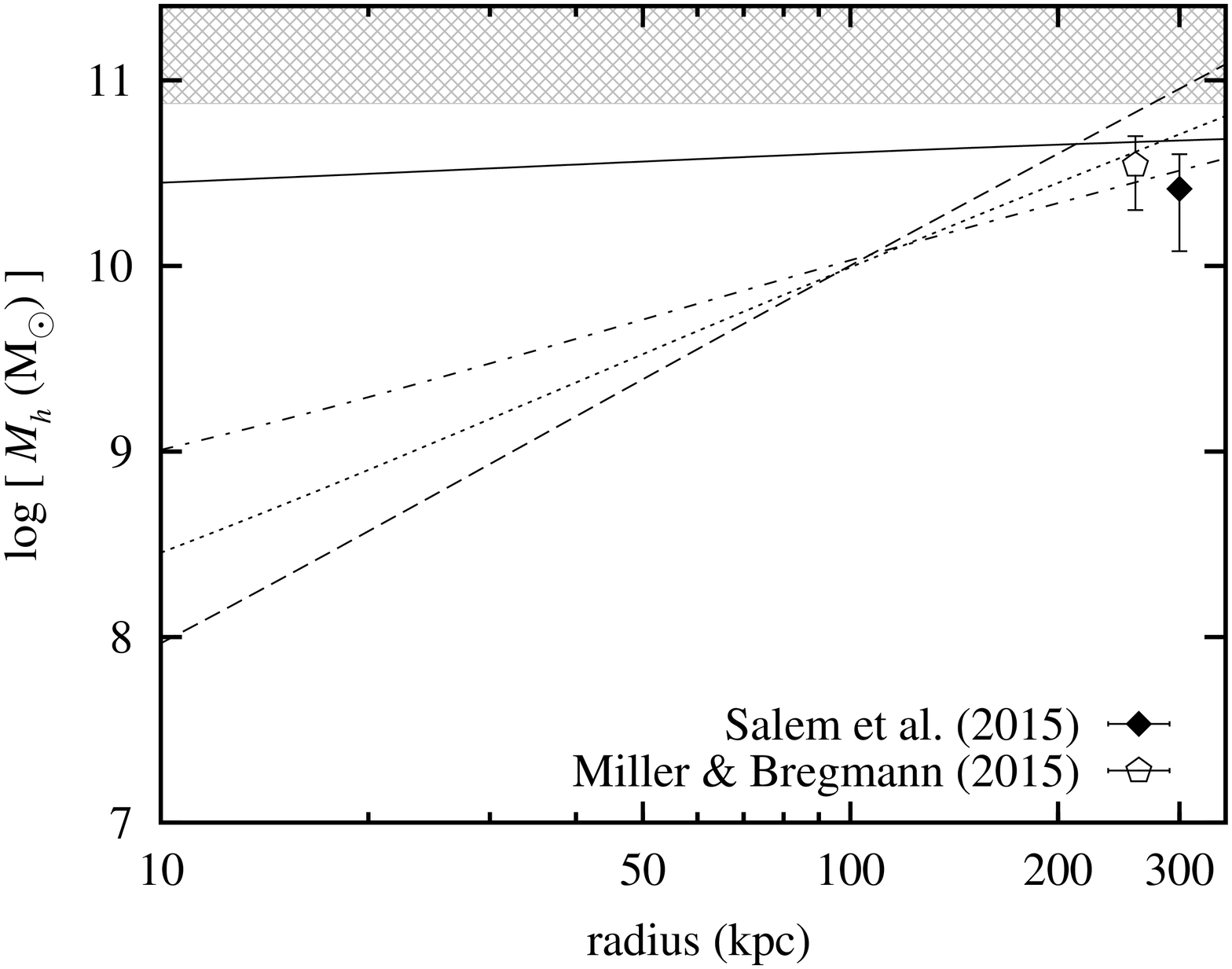}\\
\includegraphics[width=0.48\textwidth]{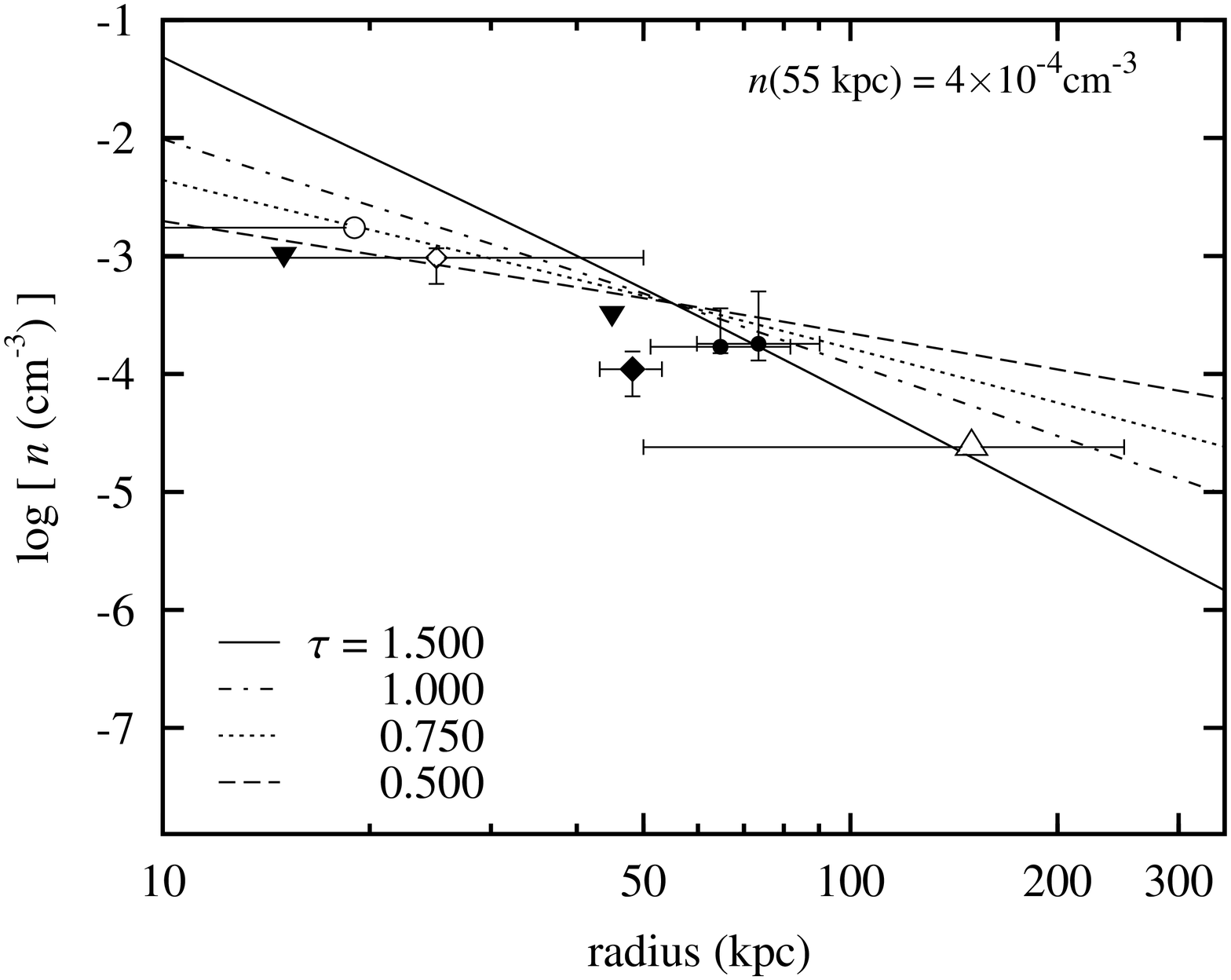}
\hfill
\includegraphics[width=0.48\textwidth]{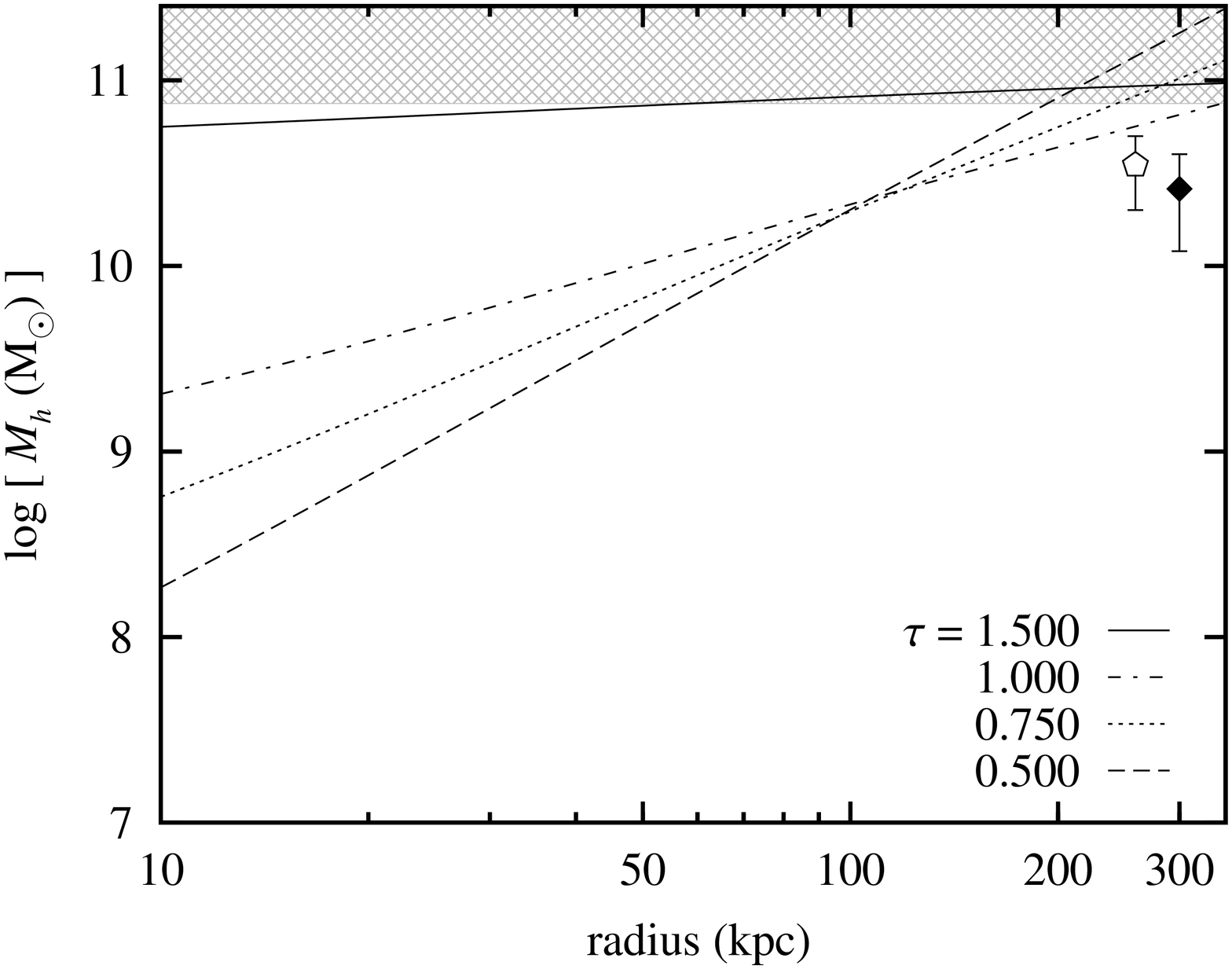}
\caption[ ]{Left: Density profile model (equation \ref{eq:partdens}) for different values of $\tratio$. Also shown are a set of halo density measurements at a range of Galactocentric distances. Right: Gas mass enclosed within a given radius (equation \ref{eq:gasm}) for each of the density profiles shown in the left panel. The data points indicate the range of Galactic halo gas masses within $r = 300 \kpc$ inferred from observations. The grey hatched area on the top of the panel indicates the range of gas masses which are too large to be consistent with the (mean) cosmic baryon-to-total mass ratio $f_b \approx 0.16$ (see text for details). Note that the top panels correspond to models that have been scaled to a fiducial value $n(55 \kpc) =  2 \times 10^{-4} \pcc$. The bottom panels show the corresponding results adopting $n(55 \kpc) =  4 \times 10^{-4} \pcc$ instead. These are discussed in Section \ref{sec:modvars}.}
\label{fig:dens}
\end{figure*}

Clearly, the density field of gas in hydrostatic equilibrium with a fixed isothermal potential will be different for different gas temperatures. If the gaseous halo is `hotter' than the DM particles, i.e. $\tratio < 1$, then the density falls off more gently with radius. Values $\tratio > 1$ on the other hand reflect the fact that the gas has cooled {\em below} the virial temperature of the DM halo, and is hence more concentrated. Large \tratio\ values thus imply lower densities in the outer region of the halo, which in turn leads to weaker hydrodynamic interactions (at a fixed distance).

Since the the virial temperature of our DM halo model is fixed, we control the temperature of the halo gas by varying \tratio. In the following we consider values of $\tratio$ in the range $[0.5, \, 1.5]$ only, and choose the set of values $\tratio \in \{1.5, \, 1.0, \, 0.75, \, 0.5 \}$ as representative of this range. These values imply gas temperatures $T_{h} \sim 10^6 \K$ (see Table  \ref{tab:gashalo}), consistent with estimates of the temperature of the Galactic hot halo \citep[][]{sno00a}. We ignore values of $\tratio < 0.5$, since these yield overly shallow density profiles, which are inconsistent with observations (see Figure \ref{fig:dens}). Similarly, we ignore values of $\tratio > 1.5$ because these imply gas temperature which are too low compared to observations (see Section \ref{sec:ic}).

We compare in Figure \ref{fig:dmhalo2} the density profiles of gas at rest in different DM potentials, all scaled to a fiducial value $n =  2 \times 10^{-4} \pcc$ at 55 kpc (see below and Section \ref{sec:ic}). For a given \tratio, the gas density profile within a NFW DM potential (as defined by the respective parameters in Table \ref{tab:dmhalo}) is steeper compared to the isothermal DM potential, but shallower than the Einasto DM halo.  The difference in the density profile across the models reflects the difference in the potential in the radial range of interest. In contrast, the gas mass enclosed within \rvir\ for all three models and a given \tratio\ are comparable (with exception of the mass for the Einasto model with \tratio\ = 0.5 which is well above $10^{11} \Msun$). Therefore, of all three models at a fixed \tratio, the isothermal sphere halo leads to {\em the highest gas density at any given distance} beyond 55 kpc for roughly the same gas mass, and hence to the {\em strongest} hydrodynamic interaction in the outer halo.

The different density profiles of gas sitting in an isothermal DM halo for our adopted values of \tratio\ (equation \ref{eq:partdens}), scaled to $n =  2 \times 10^{-4} \pcc$ at 55 kpc, are shown in the top-left panel of Figure \ref{fig:dens}.  For comparison, we include there a set of values of the halo density at various Galactocentric distances obtained from observations using a variety of methods (see Section \ref{sec:ic}). It is reassuring that all our models are fairly consistent with these measurements.

The top-right panel of Figure \ref{fig:dens} displays the mass enclosed within a given radius (equation \ref{eq:gasm}) for each of the models shown in the top-left panel. As a consistency check, we compare the model masses to the upper limit on the Galactic baryion budget within \rvir\ set by the universal mean baryon-to-total mass ratio $f_b \equiv \Omega_b / \Omega_m$. The most recent estimates of the baryon and cold dark matter mass densities $\Omega_b h^2 = 0.02205 \pm 0.00028$ and $\Omega_c h^2 = 0.1199 \pm 0.0027$ \citepalias{pla14b} imply $f_b \approx 0.16$. This value, together with the total mass of the Galaxy ($M_{tot} \sim 10^{12} \Msun$) allow for a maximum gas mass of the Galaxy within \rvir\ below $10^{11} \Msun$. Clearly, all our models result in masses within \rvir\ which are below this limit, indicated by the grey hatched area in the right panel of Figure \ref{fig:dens}. Note that a density profile shallower than the \tratio\ = 0.5 model, or a value of $n(55 \kpc)$ significantly higher than our fiducial value, would result in a gas mass largely inconsistent with these constraints. We include in this figure the range of hot halo masses inferred from observations by \citet{mil15a}, comparable to the masses estimated by others \citep[e.g][]{gat13a}, and the somewhat lower values inferred by \cite{sal15a}. Note that these estimates are all directly comparable to our model results since in all cases a smooth, monotonically decreasing density profile has been assumed. All our models, with exception of the \tratio\ = 0.5 model, predict masses within \rvir\ that are consistent with the inferred mass. 

Thus, the model \tratio\ = 0.5 appears to be marginally consistent both with the mean density of the Galactic corona at large distances, and with the constraint on the total gas mass of the Galaxy. The model \tratio\ = 1.5, although compatible with these constraints, appears too concentrated to be a plausible description of the Galaxy's halo. In contrast, the models \tratio\ = 0.75 and \tratio\ = 1 both display the best performance in terms of both the density profile and the gas mass enclosed within the virial radius of the Galaxy, although the former model yields a slightly hotter and more massive gas halo. In addition, these models reproduce by construction the relevant properties of the Galactic DM halo. Thus. we consider these models in particular provide a fully self-consistent and well founded description of the Galactic hot halo, despite its idealized nature.

\begin{table}
\caption{Isothermal gaseous halo properties}
\label{tab:gashalo}
\begin{center}
\begin{tabular}{r@{.}lcc}
\hline
\hline
\multicolumn{2}{c}{\tratio}   &   $T_{h}$  [$10^6 \K$] &   $M_{h}(\!\rvir)$ [$10^{10} \Msun$] ~\\
\hline
0&5    &	1.16    &	3.36    ~\\
0&75    &	0.77   &	2.06    ~\\
1&0    &	0.58    &	1.41    ~\\
1&5    &	0.39    &2.33    ~\\
\hline
\end{tabular}
\end{center}
\end{table}%

\section{Numerical experiment} \label{sec:exp}

We simulate the passage of a stream of gas emulating the Magellanic Stream in its orbit through the Galactic hot halo by means of a 3-dimensional (3D) `wind-tunnel experiment', expanding on the work by \citet{bla07a}. In brief, we place a warm and essentially neutral, {\em fractal} gas cloud at a distance \rms\ initially at rest with respect to the computational volume, and exposed it to a hot wind at a constant temperature $T_h$ (for a fixed \tratio) and constant density $n$ (for a fixed \tratio\ and \rms) flowing with velocity $\vec{v}_h$ under a fixed impact angle $\vartheta$ with respect to the gas cloud (see below). The warm gas is assumed to be initially in a state of pseudo-equilibrium with the hot gas, defined by the mean cloud-to-halo density ratio (or overdensity) $\eta \equiv \rho_{w} / \rho_h$, and the mean cloud-to-halo pressure ratio $\xi \equiv P_{w} / P_h$ (see equation \ref{eq:tempw}).

\subsection{Code} \label{sec:cod}

We choose for our experiment the high-resolution, multi-phase, shock-capturing hydrodynamic grid-based code \fyra\  \citep{sut10a}, especially developed for astrophysical applications. The code solves the fluid dynamic equations in one, two, and three dimensions as required. It has been shown to be fast, robust and accurate when compared to similar codes, and it performs well when subject to a standard suite of test cases as developed by \citet{lis03a}. A unique feature of the \fyra\  code is that it includes non-equilibrium cooling through time-dependent ionisation calculations. In addition, the code allows for the use of a variable equation of state (EoS) through a variable adiabatic index $\gamma$ and / or a variable mean molecular weight $\mu$. These features are essential due to the large difference in the relevant time-scales which determine the physical state of multi-phase gas, as well as the large range of densities encountered in these type of simulations.

\subsection{Observational constraints, initial conditions, and set-up}  \label{sec:ic}

The proper motion of the Large Magellanic Cloud (LMC) has recently been measured using {\em HST} data, yielding an orbital velocity $v_{\textnormal{\sc lmc}} = 321 \pm 24 \kms$ \citep{kal13a}. We adopt the high-end value and set the speed of the hot wind to $v_h = 350 \kms$. Also, we adopt a value for the impact angle $\vartheta = 24^\circ$, such that the (shear) velocity of the hot wind is given by $\vec{v}_h = (v_h \cos \vartheta, \, v_h \sin  \vartheta, \, 0)  = (320, \, 141, 0) \kms$. Note that our results are fairly insensitive to the adopted value of $\vartheta$, as long as $\vartheta \gg 0$ and $\vartheta \ll \pi / 2$, which is supported by the believe that the orbit of the Magellanic Stream is likely neither radial nor tangential with respect to the gaseous Galactic halo. We will assess the impact of this plausible, albeit arbitrary, choice on our results when dealing with virtual observations in Section \ref{sec:results} below.

The Stream's mean metallicity away from the MCs is now well constrained to $Z \approx 0.1 \Zsun$ \citep{ric13a,fox14a}. In contrast, the  metallicity of the halo gas is still uncertain, although cosmological simulations \cite[][]{ras09a} and pulsar dispersion measures towards the LMC \citep{mil15a} both suggest that it is likely in the range $Z \sim 0.1 \Zsun - 0.3 \Zsun$ far away from the disk. We choose a value for the metallicity of the halo of $Z = 0.1 \Zsun$, which is consistent with the mean value observed in external galaxies similar to the MW \citep[NGC~891;][]{hod13a}.

\begin{table}[htbp]
\caption{Relevant simulation parameters / initial conditions}
\label{tab:sims}
\begin{center}
\begin{tabular}{lll}
\hline
\hline
 Parameter				& Value					&	Remarks							~\\
\hline
$(n_x, n_y, n_z)$			& $(432, 216, 216)$		& Grid dimensions					~\\
$(x, y, z)$ [kpc]			& $(18, 9, 9)$			& Physical dimensions					~\\
$\delta x$ [pc]			& 42					& Spatial resolution (approximate)		~\\
$T_{h} ~[\!\K]$			& $10^6$				& Halo gas temperature$^a$			~\\
$T_{w} ~[\!\K]$			& $10^3$ 				& Initial Stream gas temperature$^b$	~\\
$M_{w}(\!\HI) ~[\!\Msun]$	& $10^7$				& Initial Stream neutral gas mass		~\\
$\eta$					& $100$ 					& Initial ratio of cloud : halo density		~\\
$\xi$					& $0.1$ 					& Initial ratio of cloud : halo pressure	~\\
$n(55 \kpc) ~[\!\pcc]$		& $2 \times 10^{-4}$		& Total particle density at 55 kpc		~\\
$Z_{h} ~[\!\Zsun]$			& 0.1 					& Halo gas metallicity					~\\
$Z_{w} ~[\!\Zsun]$		& 0.1 					& Stream's metallicity					~\\
$X$						& 0.7154					& Hydrogen mass fraction				~\\
$Y$						& 0.2703					& Helium mass fraction				~\\
$\vartheta ~[^{\circ}]$		& 24					& Impact angle						~\\
$\Delta v ~[\!\kms]$		& 200					& Velocity range of emission spectra	~\\
						& 						& with pixel size $\delta v = 2 \kms$.				
\end{tabular}
\end{center}
\tablecomments{$^a$ Approximate. See Table \ref{tab:gashalo}. $^b$Approximate. The exact value will vary by small factors depending on $T_h$ (see equation \ref{eq:tempw}).}
\end{table}%

Only gas clouds that are not overly dense and which have low pressure support with respect to the ambient medium will be disrupted in realistic timescales of $\sim 100 \Myr$ \citep{bla09b}. We adopt $\eta = 100$ and $\xi = 0.1$. With these parameters fixed, the initial temperature of the warm gas phase is set by the temperature of the hot gas phase through
\begin{equation} \label{eq:tempw}
	T_{w} = \left( \frac{ \mu_{w} }{ \mu_h } \right) \left( \frac{ \xi }{ \eta } \right) T_h \, ,
\end{equation}
where it should be noted that the mean molecular weight will be generally different in each phase.

A key parameter of the models is the normalisation of the density profile at the {\em canonical} distance of the Stream above the SGP ($d = 55 \kpc$). Although still uncertain, different lines of evidence indicate that it is likely in the range of $10^{-5} \pcc - 10^{-3} \pcc$ at $20 \kpc\ \lesssim r \lesssim 100 \kpc$. For example, \citet{bli00a} estimate a lower limit on the mean halo density of $n \approx 2.4 \times 10^{-5} \pcc$ out to $d \leq 250 ~\kpc$ based on the assumption that the gas-poor dwarf spheroidals orbiting the Galaxy have been stripped from their gas by the ram-pressure exerted by the hot halo. Along the same line, and combining observations with (2D) hydrodynamic simulations, \citet{gat13a} have inferred a range of halo densities $n \approx (1 - 4) \times 10^{-4} \pcc$ at $50 ~\kpc < d < 100 ~\kpc$. \citet{sta02a} have found that the gas clouds at the tail of the MS are likely in pressure equilibrium with the hot halo, and using this they have put an upper limit on the halo density of $10^{-3} \pcc$ and $3 \times 10^{-4} \pcc$ at a distance $z = 15 ~\kpc$ and $z = 45 ~\kpc$ from the Galactic plane, respectively. \citet{and10a} infer a range for the mean halo density of $n \approx (6 - 10) \times 10^{-4}$ out to the LMC ($d \approx 50 \kpc$) based on dispersion measures of LMC pulsars. However, the stripping of the LMC's disc requires a somewhat lower value of $n(48.2 \pm 3 \kpc) = (1.1 \pm 0.44) \times 10^{-4} \pcc$ \citep{sal15a}.

Based on these results, we adopt a fiducial value $n(55 \kpc) =  2 \times 10^{-4} \pcc$, consistent with \citet{bla07a}. As shown in Figure \ref{fig:dens}, this choice leads to models for the Galactic corona that largely agree with the results from observations over a broad range in distances. In this respect, we consider both $n(55 \kpc)$ and \tratio\ to be well constrained by observation. Note that the models are completely defined by the value of $\tau$, given that all the other parameters are either fixed or they depend on $\tau$ (Table \ref{tab:sims}).\\

We run all simulations in a rectangular box of comoving size $18 \times 9 \times 9 \kpc^3$, using a fixed grid composed of $432 \times 216 \times 216$ cells. These settings imply a spatial resolution of ${\delta x}= ( 9 / 216) \kpc\ \approx 42 \pc$. The fragment of gas representing the Magellanic Stream is initially constrained to a cylinder $18 \kpc$ in length and $2 \kpc$ in diameter, and whose axis of symmetry runs parallel to the $x$ axis of the coordinate system defined by the box. The impact angle is defined with respect to the $x$-axis of this cylinder. In this setup, the $x$-axis coincides with the Magellanic longitude, $l_M$, whereas any of $y$ or $z$ run along the Magellanic latitude, $b_M$ \citep[][]{wak01a,nid08a}. The simulated Magellanic Stream consists of an \HI\ gas distribution initially at temperature $T_{w} \sim 10^3 \K$ (see equation \ref{eq:tempw}); a mean initial hydrogen particle density $n \sim 10^{-3} \pcc$; and a total neutral gas mass $M_{w}(\!\HI) \sim 10^7 \Msun$. The initial warm gas density field corresponds to the density of a fractal medium described by a Kolmogorov turbulent power spectrum $P(k) \propto k^{-5/3}$, with a minimum wavenumber $k_{min} = 8$ (relative to the grid) corresponding to a spatial scale of 2.25 \kpc\ \citep[q.v.][]{sut07a}, comparable to the typical size of clouds in the Stream. The \HI\ gas cloud is assumed to be at a fixed distance \rms\ from the Galactic Centre in the direction of the SGP. The temperature $T_h$ and the density $n$ of the hot wind are set according to the value of $\tau$ and the distance \rms\ as given by equations \eqref{eq:tdm} and \eqref{eq:partdens}, respectively.

For each $\tau \in \{0.5, \, 0.75, \, 1.0, \, 1.5  \}$, we consider a set of Galactocentric distances $r_{\textnormal{\sc ms}} \in \{55, \, 75, \, 100, \, 125, \, 150 \} \kpc$ which together span the range of plausible orbits of the MS above the SGP \citep{gug14a}. This yields 20 models. Each model is run for a total (simulation) time of $320 \Myr$, starting from $t_{sim} = 0 \Myr$, assuming free boundary conditions. The simulation output for a given set of values $\{\tau, \, \rms, \, t_{sim} \}$ -- in steps of $\Delta t_{sim} = 10 \Myr$ -- consists of a series of datacubes containing information about the \HI- and \HII\ densities, $n_{ \HI }$ and $n_{\!\HII }$, respectively; the gas temperature $T$, and the gas velocity $\vec{v} = (v_x, \, v_y, \, v_z)$. Using this information, we compute the \Ha\ emission and the \HI\ column density of each cell, and from these the \Ha\ surface brightness and total \HI\ column density along the \los\ for each snapshot.

\subsection{Emission line spectra}

We compare the \Ha\ emission of the gas in our simulations to Fabry-P\'erot \Ha\ observations along the Magellanic Stream \citep[][and references therein]{bla13a}, and complement these with results on the associated \HI\ column density measurements.

Given the one-to-one correspondence between the \HI\ 21cm emission (i.e., brightness temperature) $T_B$ and the \HI\ column density of a parcel of optically thin gas \citep[e.g.][]{dic90b},
we use the total \HI\ column density along the \los\ (\NHI) as a proxy for the corresponding \HI\ 21cm emission, i.e. we {\em define} the \HI\ 21cm intensity to be $I_{\!\HI} \equiv \NHI$.

The \Ha\ emission is computed using
\begin{equation} \label{eq:mua}
	\mua =  \mua^{(shock)}  + \mua^{(phot)} \, .
\end{equation}
The first term accounts for the ionisation that results from slow shocks produced by the collision of the trailing cloud gas with the leading gas ablated by the interaction with the hot halo, and is given by
\begin{equation} \label{eq:hacell}
	\mua^{(shock)}  = [1 + (Y / 4 X)] ~K_{R} \; \alphabha \; \!\! \int \! (n_{\rm HII})^2 ~ds\, ,
\end{equation}
where $\alphabha(T)$ is the effective \Ha\ recombination coefficient (equation \ref{eq:peqeff}), $K_R  \approx 1.67 \times 10^{-4}  \cm^2 ~{\rm s} \mR$,\footnote{1 milli-Rayleigh (mR) corresponds to \mbox{ $10^3 / 4 \pi$ photons $\psc \ps \psr$} \citep{bak76a}, or $2.41 \times 10^{-4} \erg \psc \ps \psr$ at \Ha.} and the  factor $[1 + (Y / 4 X)]$ accounts for the conversion of electron density to \HII\ particle density. Adopting a hydrogen and helium mass fractions $X = 0.7154$ and $Y = 0.2703$, respectively \citep[][solar bulk composition]{asp09a},\footnote{For our adopted metallicity of 10 percent the solar value, $\Zsun = 0.0142$, the contribution of heavy elements to the electron density can be neglected.} assuming the gas is fully ionised ($\!\nH \approx n_{\!\HII}$), it follows that $[1 + (Y / 4 X)] \approx 1.09$.\footnote{We assume that helium is only singly ionised, given that the ionisation energy of \ion{He}{\rm II} is comparatively high \citep[$E \approx 54.4 ~{\rm eV}$;][]{kra14a}.}

The second term in equation \eqref{eq:mua} accounts for the ionising effect of the cosmic ultraviolet background radiation (UVB). The \Ha\ emission along the \los\ of gas in photoionisation equilibrium with the UVB radiation field is
\begin{equation} \label{eq:haphot}
	\mua^{(phot)} = \frac{ 1 }{ 4 \pi } \Gamma_{\!\HI} \; \!\! \int \! (f_{\!\Ha} \; n_{\!\HI}) ~ds \, .
\end{equation}
where $f_{\!\Ha}(T)$ gives the fraction of recombinations that produce an \Ha\ photon, and $f_{\!\Ha}(10^4 \K) \approx 0.45$ (equation \ref{eq:fha}). The \HI\ recombination rate, $\Gamma_{\!\HI}$, is related to the total ionising photon flux $\Phi_i$ through
\begin{equation} \label{eq:ionflux}
	\Phi_i = 1.59 \times 10^4 ~{\rm photon} \psc \ps ~\left( \frac{\gamma + 3}{4 \gamma} \right) ~\left( \frac{ \Gamma_{\!\HI} }{ 10^{-13} \ps} \right) \, .
\end{equation}
We adopt $\Gamma_{\!\HI} = 10^{-13} \ps$ \citep[appropriate for $z = 0$;][]{wey01a} and $\gamma = 1.8$ \citep[][]{shu99b}, corresponding to an ionising flux $\Phi_i \sim 10^4$ photons \psc\ \ps. If we used instead the most recent estimate $\Gamma_{\!\HI} = 4.6 \times 10^{-14} \ps$ \citep{shu15a}, the flux would be lower by roughly a factor 2.

To mimic radiation transfer effects, we limit the depth (along the \los\ in any direction) of the gas ionised by the UVB to a maximum value defined by the condition that the column recombination ($n_e \, n_{\!\HII} \, \alpha_B$) equals the incident ionising photon flux ($\Phi_i$). This condition is equivalent to restricting the ionising effect of the UVB to a column of neutral gas $\sim 10^{17} \psc$ (see Appendix \ref{sec:pi}). The effect of the cosmic UVB is to produce an ionisation skin around the cloud featuring an \Ha\ surface brightness at a level of roughly 5 mR. Again, if we used instead $\Gamma_{\!\HI} = 4.6 \times 10^{-14} \ps$, this value would decrease to roughly 2 mR. It is important to mention that this approach is not entirely self-consistent with our simulations because the ionising effect of the UVB is not included at runtime, and because it assumes photoionisation equilibrium. Also, we ignore for the moment the contribution of the Galactic ionising field, which would produce an additional mean \Ha\ signal of $21 \, \zeta  \, (d / 55 \kpc)^{-2} \mR$ \citep[$\zeta \approx 2$;][]{bla13a}. 

\begin{figure*}[htpb]
\includegraphics[width = 0.5\textwidth]{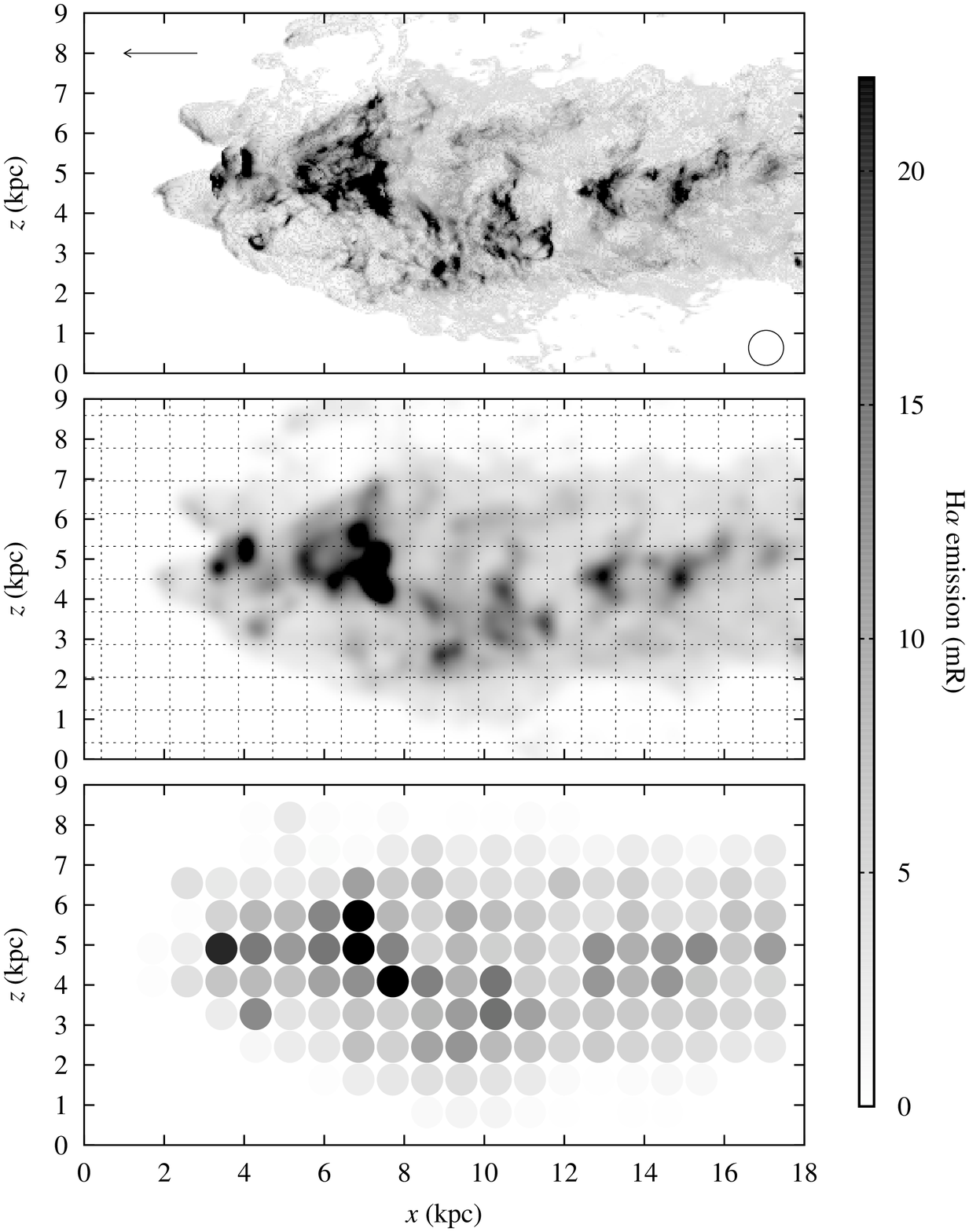}
\includegraphics[width = 0.5\textwidth]{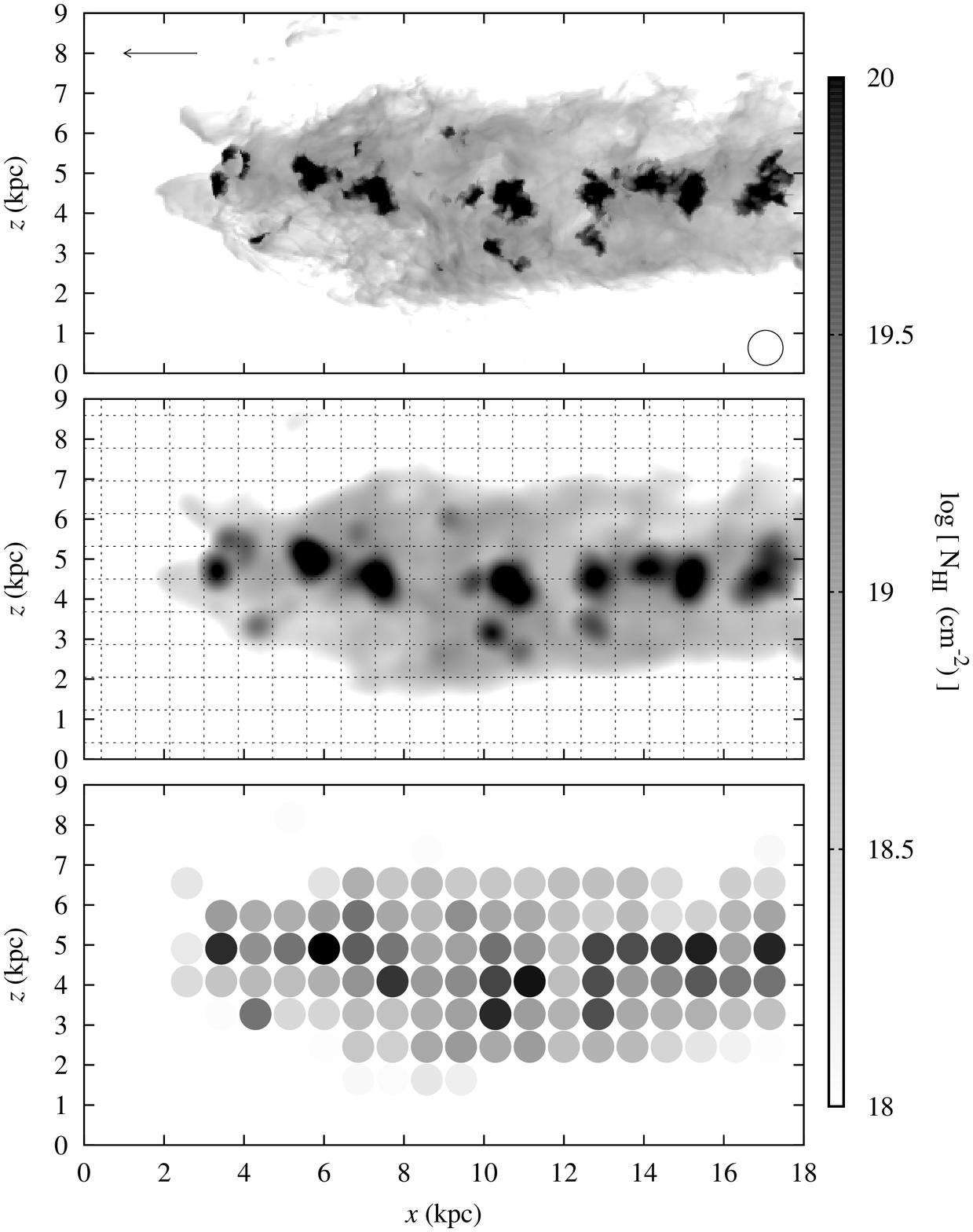}
\caption[\!\Ha\ / \HI\ 21cm emission along the Magellanic Stream ($y$ projection)]{Virtual observation of the \Ha\ (left) and \HI\ 21cm (right) emission along the Stream in model \tratio\ = 1 and \rms\ = 55 \kpc\ at $t_{sim} = 170 \Myr$, projected along the $y$ axis. Top: Emission at the intrinsic (high) resolution of the simulation. The arrow on the top-left corner indicates the projected Stream's velocity. Middle: The result of spatially smoothing the emission map using a circular Gaussian kernel of size $0.5^{\circ}$ / \onedeg\ (FHWM /  full width), as indicated by the circle on the bottom-right corner of the top panel. Bottom: Emission at different pointings obtained using a beam with an angular diameter of \onedeg\ on the sky. The position of each individual pointing is defined by each cell of the grid overplotted on the middle panel. The grayscale bar to the right of each set of panels indicates the \Ha\ intensity in milli-Rayleigh (left) and the \HI\ column density in $\psc$ (right). The full evolution of the \Ha\ surface brightness and the \HI\ column density are available at \url{http://www.physics.usyd.edu.au/~tepper/proj_ms.html}.}
\label{fig:hamap}
\end{figure*}

To allow for a faithful comparison with observations, we map the simulation data onto observed space by projecting the simulation volume along a given axis, so as to mimic the projection of the observed emission along the Stream onto the plane of the sky. We choose, for convenience, an axis parallel to one side of the simulation box, and perpendicular to the Stream's main axis, i.e. the $y$-axis. However, we will address the potential bias introduced by this choice by comparing the results obtained by projecting along all three orthogonal axes, $x$, $y$ and $z$.

Each \los\ across the projected datacube thus corresponds to a pencil-beam spectrum. \Ha\ and \HI\ 21cm pencil-beam spectra along the chosen projection axis are constructed by computing the intensity for each cell in the simulation volume. The \Ha\ intensity and the \HI\ 21cm emission intensity observed at velocity $v_m$ (in the rest-frame of the \HI\ gas at $t_{sim} = 0$) of a parcel of gas at cell $\mathbf{n} \equiv (i, j, k)$ with bulk velocity $v_{\mathbf{n}}$ are, respectively,
\begin{align} \label{eq:spec}
	I_{ \textnormal{\Ha}} (v_m - v_{\mathbf{n}}, \mathbf{n}) & =  \mua(\mathbf{n}) \; \phi_{\mathbf{n}}(v_m - v_{\mathbf{n}}) \, , \notag \\
	I_{\!\HI} (v_m - v_{\mathbf{n}}, \mathbf{n}) & = \NHI(\mathbf{n}) \; \phi_{\mathbf{n}}(v_m - v_{\mathbf{n}}) \, ,
\end{align}
where the normalised line profile \mbox{$\phi(v) = \exp{ [ - v^2 / b_T^2 ] }$}, with \mbox{$b_T= ( 2 k T_{w} / m_u )^{1/2}$}.

We adopt a spectral range in terms of velocity of $\Delta v = 200 \kms$ which corresponds to the spectral range provided by the WHAM spectrometer. The velocity scale is given with respect of the initial rest-frame of the \HI\ gas, such that emission spectrum spans the range $[-100, +100] \kms$, with a pixel size $\delta v = 2 \kms$.

At each beam position (or `pointing'), we compute the intensity-weighted average of all cells within the beam, resulting in a single spectrum per pointing. We adopt a low resolution beam of $\onedeg$ diameter on the sky, identical to the resolution provided by the WHAM spectrometer.

We choose a rectangular (rather than a circular) beam, which allows for a full coverage of the projected image, and which greatly simplifies the scanning procedure.\footnote{A alternative arrangement consisting of a tightly packed array of circular windows \citep[which conveys equal weight to every pixel within the beam; see e.g.,][ their Figure 3]{haf03a} would yield essentially the same results.} Given that a beam with a diameter of $2 \theta$ subtends a solid angle $\Omega_{beam} \approx \pi \left( \theta / 2 \right)^2$ (provided that $\theta \ll 1$), the linear size of a square subtending a solid angle $\Omega_{beam}$ at a distance \rms\ relative to the angular dimension of a single cell in our simulation, $\delta \Omega_V  \approx ( {\delta x}/ \rms )^2$, is roughly ${\delta l} = (\pi / 4 )^{1/2} (\!\rms / \delta x ) ~\theta$. The emission within a beam pointing at each velocity $v_m$, i.e. the beam spectrum, is then
$$
	\overline{ I }_{ \textnormal{X} } = [{\delta l}]^{-2} \sum_{[{\delta l}]}  \sum_{[{\delta l}]} I_{ \textnormal{X} } \, ,
$$
where the sum extends over all cells within $\Omega_{beam}$, and $X \in \{\!\HI, \HII \}$. Here, the notation $[k]$ indicates the largest odd integer smaller than or equal to $k$.

The average of the emission within each beam pointing can effectively be obtained by overlaying a rectangular grid on the projected image with a cell size equal the solid angle subtended by a circular beam of diameter $2 \theta$ at that distance, and computing the arithmetic mean within each new cell. We choose the origin of the matrix to be shifted by half a beam size in each direction ($xy$) to avoid the uncertainties associated with the simulation volume's boundaries.

It is important to emphasise that the pixelation of the simulation volume and the position of the beam pointings with respect to the projected datacube are rather arbitrary. Also, the latter is also generally different for each adopted distance \rms. This results from the fact that for a computational volume with fixed comoving size and fixed grid, and a beam of fixed angular size, an increasing fraction of the gas that represents the Stream will be sampled by the beam with increasing distance.
Because of this, and also to avoid a bias in the resulting emission introduced by potentially bright features induced by chance alignments (rather than due to intrinsically bright gas blobs), {\em prior} to computing the spatial average within each beam pointing we smooth each 2D spatial slice at each velocity bin using a circular Gaussian kernel with a full width at half maximum of half the beam size and a total width of (i.e., truncated at) the size of the beam. According to the approach described above, the projected intrinsic \Ha\ emission map from our simulation is smoothed at each given distance using a Gaussian kernel with FWHM of $[{\delta l} / 2]$ pixel and a total width of $[{\delta l}]$ pixel.

Finally, in order to take into account the instrumental line broadening, we convolve each spatially averaged spectrum using a Gaussian kernel\footnote{Note that the LSF of the WHAM spectrometer is only poorly approximated by a Gaussian \citep[][]{tuf97a}. However,  given the typical \Ha\ / \HI\ 21cm line widths (20 \kms\ - 40 \kms), this approximation hardly affects our results.} with a \mbox{FWHM = 12$\kms$}, which roughly corresponds to the resolution of the WHAM spectrograph \citep[][]{rey98a}. The \Ha\ surface brightness and \HI\ 21cm emission maps are simply obtained from the 3D spectral datacube by simply integrating each spectrum along the velocity coordinate (equations \ref{eq:spec}).

A selected example illustrating the result of the above procedure is shown in Figure \ref{fig:hamap} (see also Figure \ref{fig:kin}). As we show below (see Section \ref{sec:ion}; Figure \ref{fig:ionfracevol}), the \HII\ to \HI\ mass ratio in this snapshot roughly matches to the corresponding ratio observed in the Magellanic Stream \citep[$\sim3$;][]{fox14a}. The effect of the ionising cosmic UVB is apparent: all the Stream gas is lit up and emitting at level of $\sim 5 \mR$. But there are brighter spots which are a consequence the shock cascade, whereby the trailing clouds collide with the material ablated by hydrodynamic instabilities from the leading gas, thus being shock ionised \citep[q.v.][]{bla07a}. Interestingly, while the bright \Ha\ spots seem to closely track the high \NHI\ parcels of gas, the converse is not true, with high \NHI\ appearing with no correspondent strong \Ha\ emission. However, these differences become less apparent, although they remain, as a result of the beam smearing. The most dramatic effect of the latter is the dilution of the \Ha\ and \HI\ signals with respect to the brightest levels seen at the (intrinsic) resolution of the simulation by a significant factor. Only the brightest spots in \Ha\ would observable with a Fabry-P\'erot interferometer for reasonable integration times, leading to a significant fraction of the ionised gas mass falling below the detection threshold.

\begin{figure*}
\centering
\includegraphics[width = 0.48\textwidth]{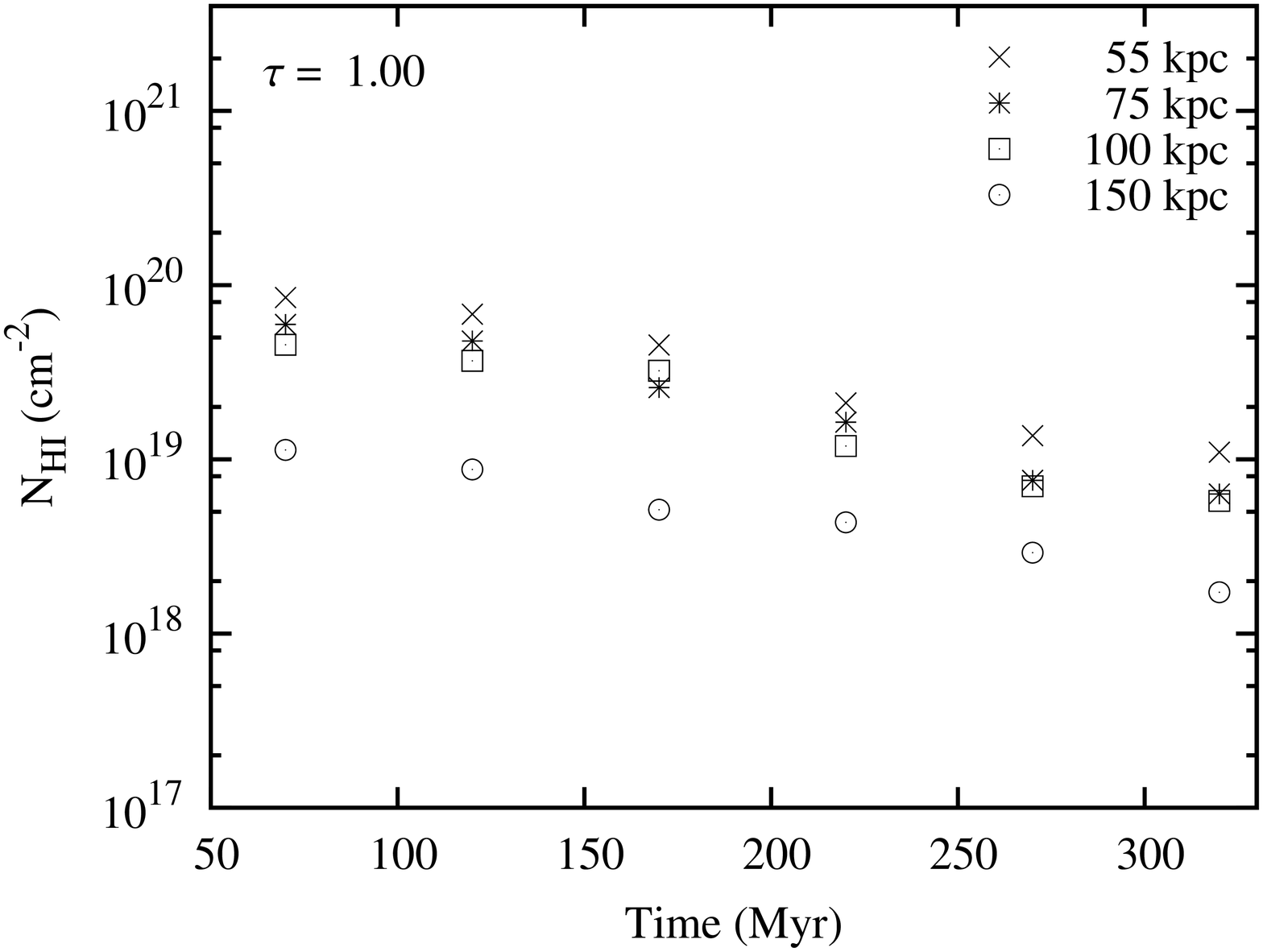}
\includegraphics[width = 0.48\textwidth]{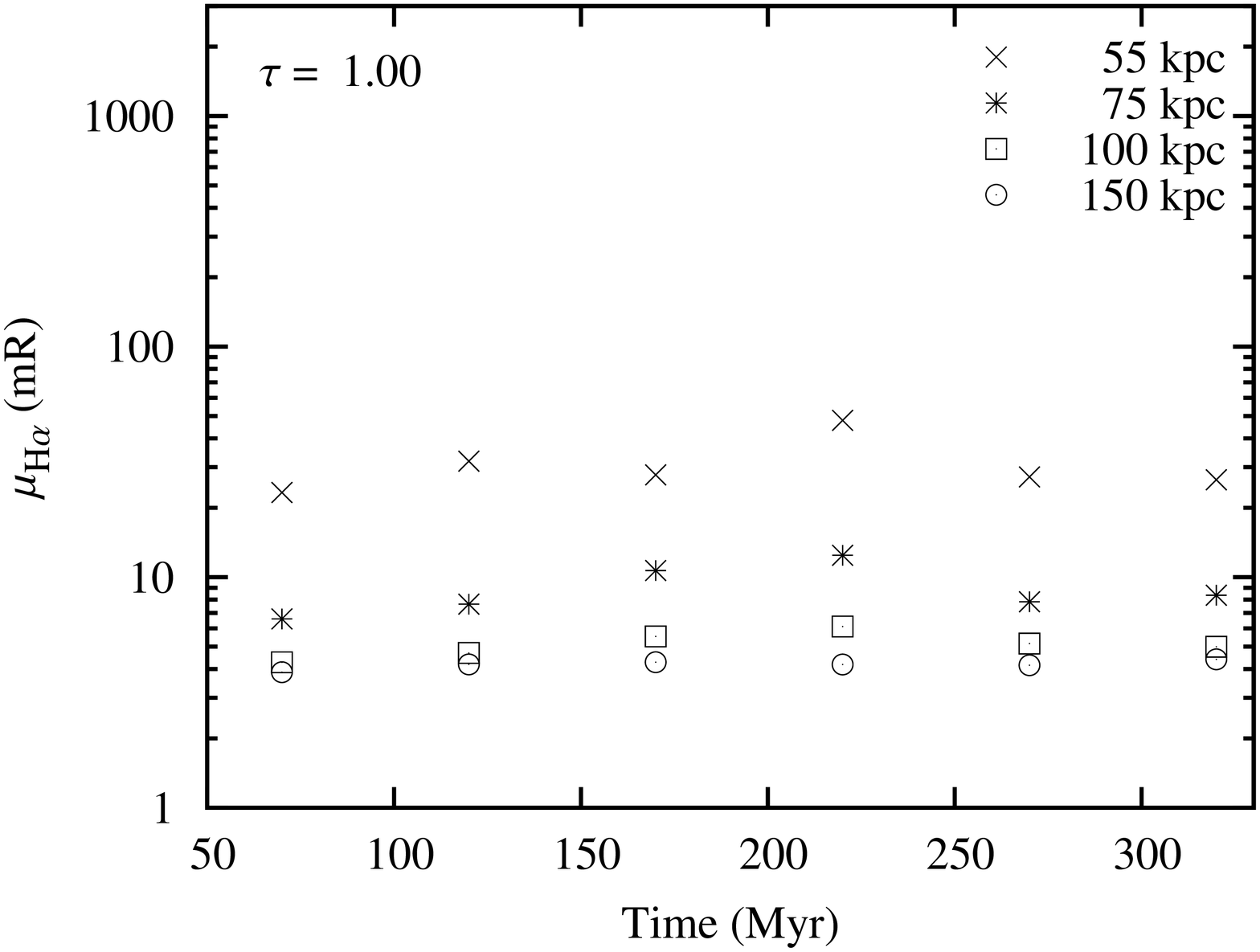} \\
\includegraphics[width = 0.48\textwidth]{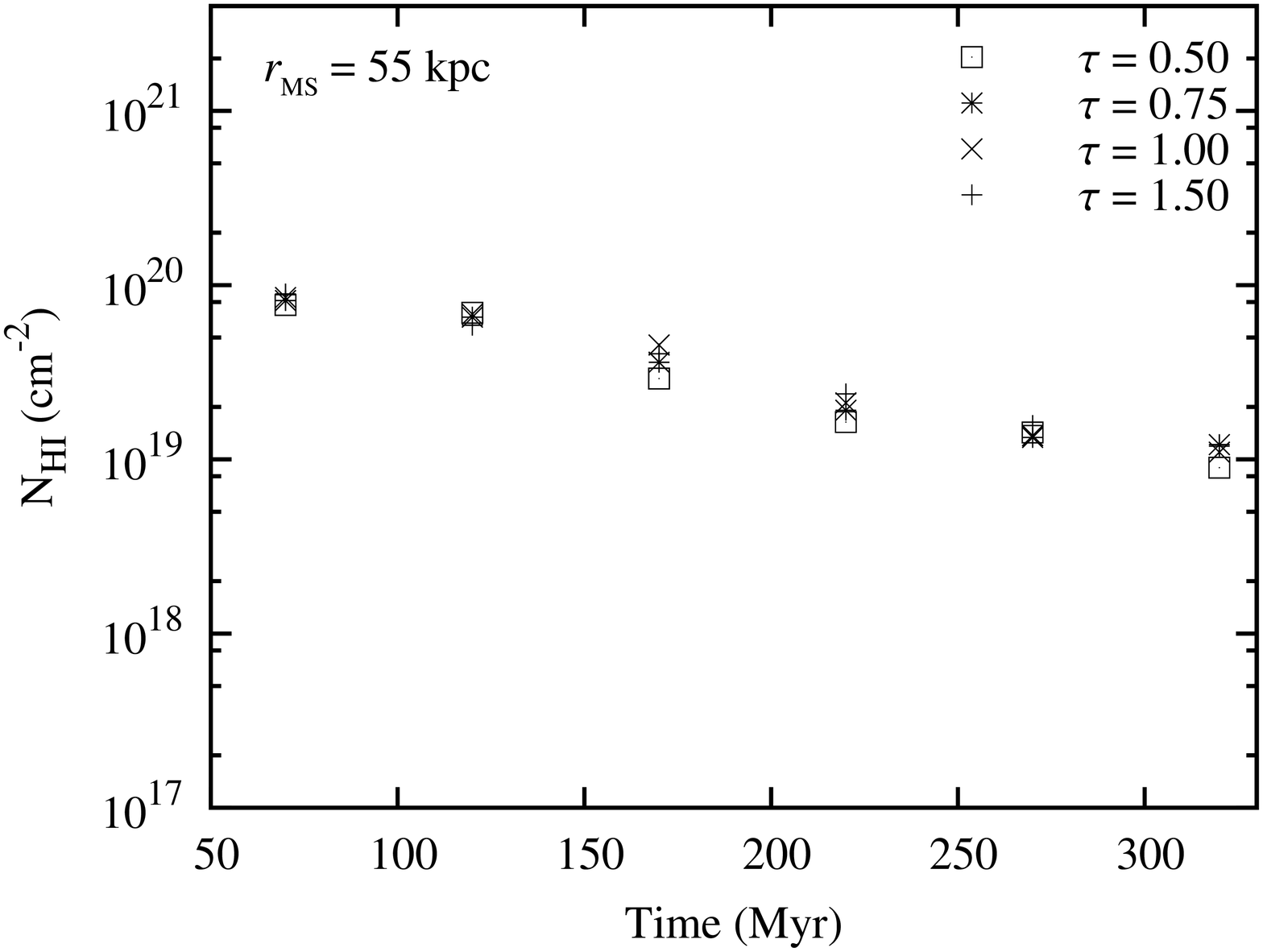}
\includegraphics[width = 0.48\textwidth]{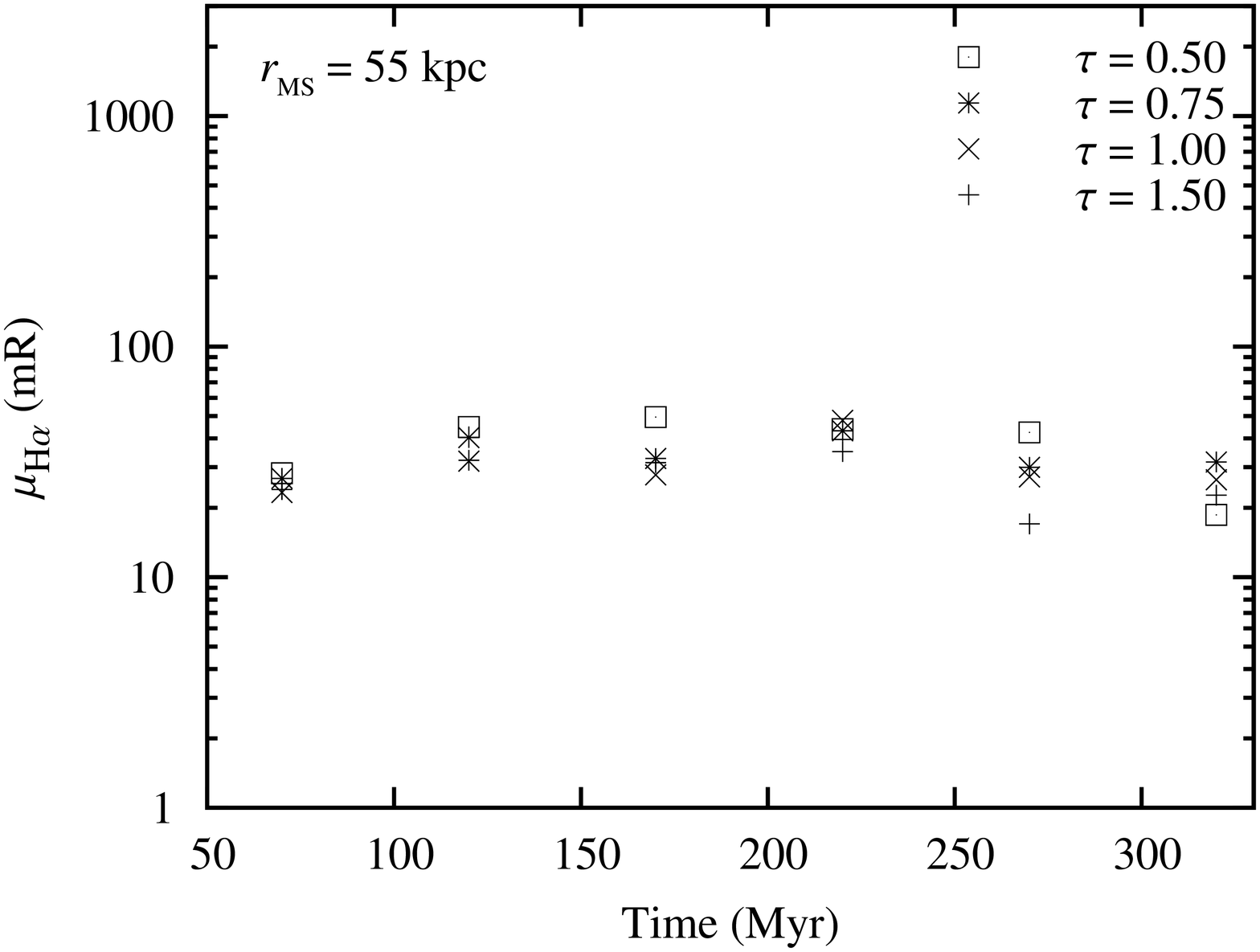} \\
\includegraphics[width = 0.48\textwidth]{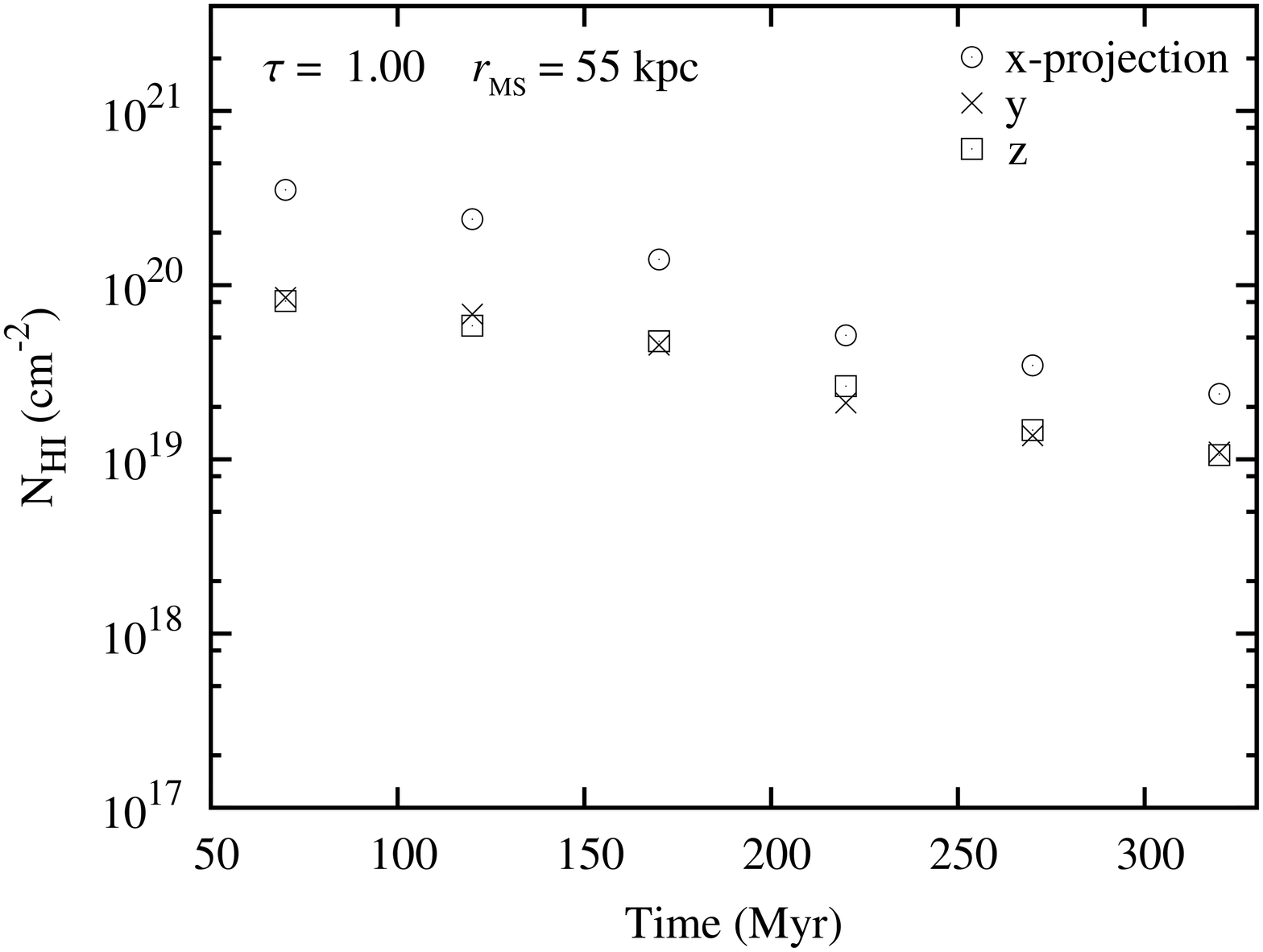}
\includegraphics[width = 0.48\textwidth]{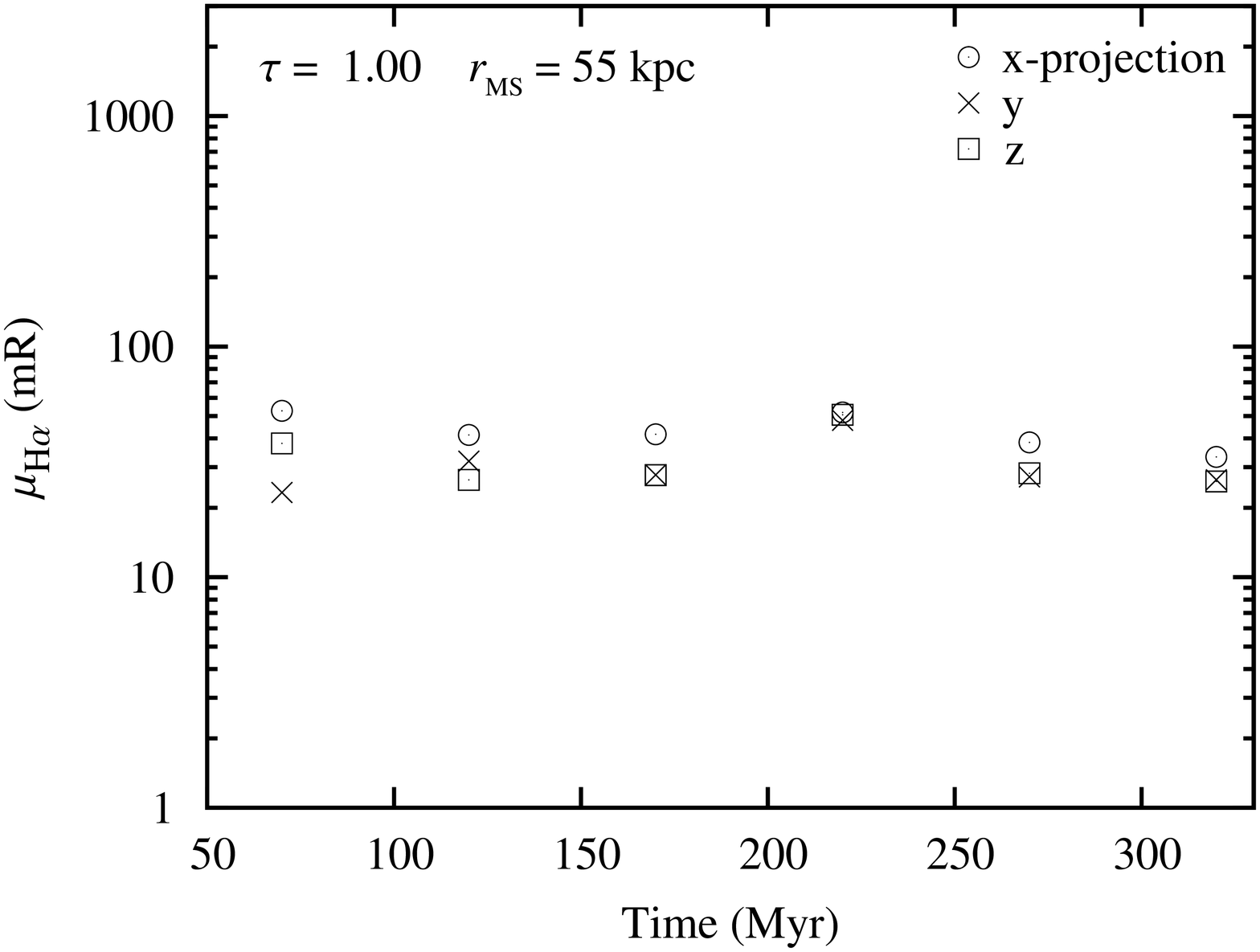}
\caption[]{Evolution of the \HI\ column density (left) and the \Ha\ surface brightness (right). The top panels correspond to the results for model \tratio\ = 1, and a range in Galactocentric distance \rms\ (indicated by the legend). The middle panels show the result at a fixed Galactocentric distance \rms\ = 55 kpc, for different \tratio\ (indicated by the legend). Each data point in the left (right) panels corresponds to the 90 (100) percentile of the distribution of intensities at that particular time and distance. The four top panels correspond to the results obtained from projecting the simulation cube along the $y$-axis. The two bottom panels illustrate the difference between projections along all three orthogonal axes for model \tratio\ = 1 at 55 kpc.}
\label{fig:haevol}
\end{figure*}

\section{Results} \label{sec:results}

\subsection{Gas emission}

We follow the evolution of the gas emission over a period of 320 \Myr\ in all 20 models. To illustrate these results, we adopt the model \tratio\ = 1 as our standard model. The top panels of Figure \ref{fig:haevol} shows the evolution of the \Ha\ / \HI\ emission at four representative distances \rms\ = 55 \kpc, 75 \kpc, 100 \kpc, and 150 \kpc. The middle panels show the corresponding result for all models \tratio\ = 0.5, 0.75, 1.0, and \tratio\ = 1.5, at a fixed distance \rms\ = 55 kpc. Note that the comparison of the results for a fixed \tratio\ at different distances allows to assess the effect of the density on the \Ha\ / \HI\ emission for a fixed temperature, while the comparison of models with different \tratio\ at a fixed distance of 55 kpc (and thus a fixed density) helps us explore the effect of the temperature.

We find that the gas ionises quickly ($\sim 50 \Myr$), and after $\sim 300 \Myr$, the \HI\ column density has decreased uniformly both with time, dropping by nearly an order of magnitude. This can be understood as a consequence of the increasing ionisation of the gas due to the interaction with the hot halo gas. The {\em detected} column density of the gas farther out is also lower with respect to the {\em intrinsic} gas density, which is an effect of the beam dilution. Note that a beam of \onedeg\ diameter samples a region of roughly twice (three times) the size at \rms\ = 100 kpc (150 \kpc) with respect to \rms\ = 55 kpc. 

Similarly, the highest \Ha\ emission -- this is, the {\em maximum} value at each given time -- comes from the gas which is closest. In contrast to the behaviour of the \HI\ density, in this case the effect is governed by the increase of halo gas density with distance, and the corresponding strength of the hydrodynamic interaction leading to the shock cascade. What is  surprising is that even at the lowest Galactocentric distance of \rms\ = 55 kpc, the emission never exceeds $\sim 40 \mR$; and it barely reaches 10 mR at 75 kpc. At even larger distances, $\rms\ \gtrsim 100 \kpc$, the emission is dominated by the recombination of the gas ionized by the cosmic UV background. Note that often we do find in our simulation pixels with $\mua > 30 \mR$, and occasionally on the order of $\sim 100 \mR$, but their strong signal is washed out as a result of the beam smearing (Figure \ref{fig:hamap}; see also Section \ref{sec:modvars}). The similarity in \Ha\ emission across models with different \tratio\ shows that these results are insensitive to variations in the halo gas temperature by factors of a few. This indicates that the \Ha\ emission is dominated by the gas ionised through cloud-cloud collisions that trigger the shock cascade, rather than the gas ionised by the interaction with the hot halo. The insignificance of the halo gas temperature together with the fact that the virial temperature across the DM halo models presented previously is very similar (Table \ref{tab:dmhalo}) makes the choice of the DM halo model irrelevant, as long as the corresponding gas density profiles are comparable.

We see that the maximum level of \Ha\ emission at any reasonable distance is comparable to, or even less than, the emission induced by the Galactic ionising starlight $\sim 20 - 40 \mR \; [d / 55 \kpc]^{-2}$. This is the reason for us to ignore this component in our models, since it would otherwise outshine the emission produced by the shock cascade. Taking the contribution of the Galactic UV into account would elevate the emission at 55 kpc to $ \sim 70 \mR$, and to $ \sim 30 \mR$ at 75 kpc, which are significantly lower than the Stream's emission observed over the SGP and at the tail of the Stream at $l_M \approx 260^{\circ}$, respectively \citep[][]{bla13a}. It is however  unclear at this point how the ionising effect of the Galactic UV included self-consistently at runtime would affect these limits.

Given that observations performed with the WHAM instrument typically reach a sensitivity of $\gtrsim 30 \mR$, the Stream gas in our models would be essentially undetectable (ignoring for the moment the contribution which results from the Galactic ionising field). In contrast, and considering that the sensitivity of e.g the GASS survey is roughly $\NHI = 1.6 \times 10^{18} \psc$, the gas at distances $d \lesssim 100 ~\kpc$ would be bright in \HI\ 21cm, and marginally detectable at $d \sim 150 ~\kpc$, even after 300 Myr.

We find that much of the gas dislodged from the main body of our model Stream is low density material that mixes rapidly with the halo gas, thereby being heated (and thus ionised) to temperatures well above $10^5 \K$, which are on the order of the temperature expected for turbulent mixing \citep{beg90a}. At these temperatures, the \Ha\ emissivity drops by nearly two orders of magnitude with respect to its value $10^4 \K$ (see equation \ref{eq:peqeff}), and the ionised gas becomes thus practically invisible in \Ha. Only the gas ionised by cloud-cloud collisions remains at relatively low temperatures ($\sim 10^4 \K$), and recombines quickly, thus providing the strongest \Ha\ signal. However, the fraction of warm ionised gas is very low overall, and thus is the corresponding \Ha\ signal. Hence, only a mechanism such as slow shocks which is able to ionise a significant fraction of the gas without increasing its temperature far above $10^4 \K$ will lead to significant \Ha\ emission.\\

As the reader may recall, all the above results correspond to virtual observations where the simulation cube has been projected along the $y$-axis. Given that this choice is somewhat arbitrary, we calculate the corresponding results for projections along all three orthogonal axes for model \tratio\ = 1 at 55 kpc. The outcome of this exercise is summarized in the bottom panels of Figure \ref{fig:haevol}. We have checked that the results are essentially the same for all other models. The \HI\ column density is consistently highest when observed along the $x$-axis, given that this axis coincides with the axis of symmetry of the initially cylindrical gas configuration in our setup, and the \los\ traverses a larger path across the \HI\ cloud. The projections along the $y$ and $z$ axes yield nearly identical results, as expected from the symmetry of the initial cloud structure.

In our experiment, the projection along the $x$-axis is equivalent to observe the Stream 'face-on'. Hence, one would na\"ively expect that the \Ha\ emission should be highest when projecting along this direction. But surprisingly, the \Ha\ intensity in our models is very similar regardless of the projection, being only slightly stronger when viewing the gas cloud face-on. This implies that the choice of projection axis to compute virtual observations is essentially irrelevant. Moreover, since different projections effectively imply significant variations in the impact angle, our choice of a particular value for $\vartheta$ turns to be irrelevant as well, as far as the \Ha\ intensities are concerned.

\begin{figure*}
\centering
\includegraphics[width = 0.48\textwidth]{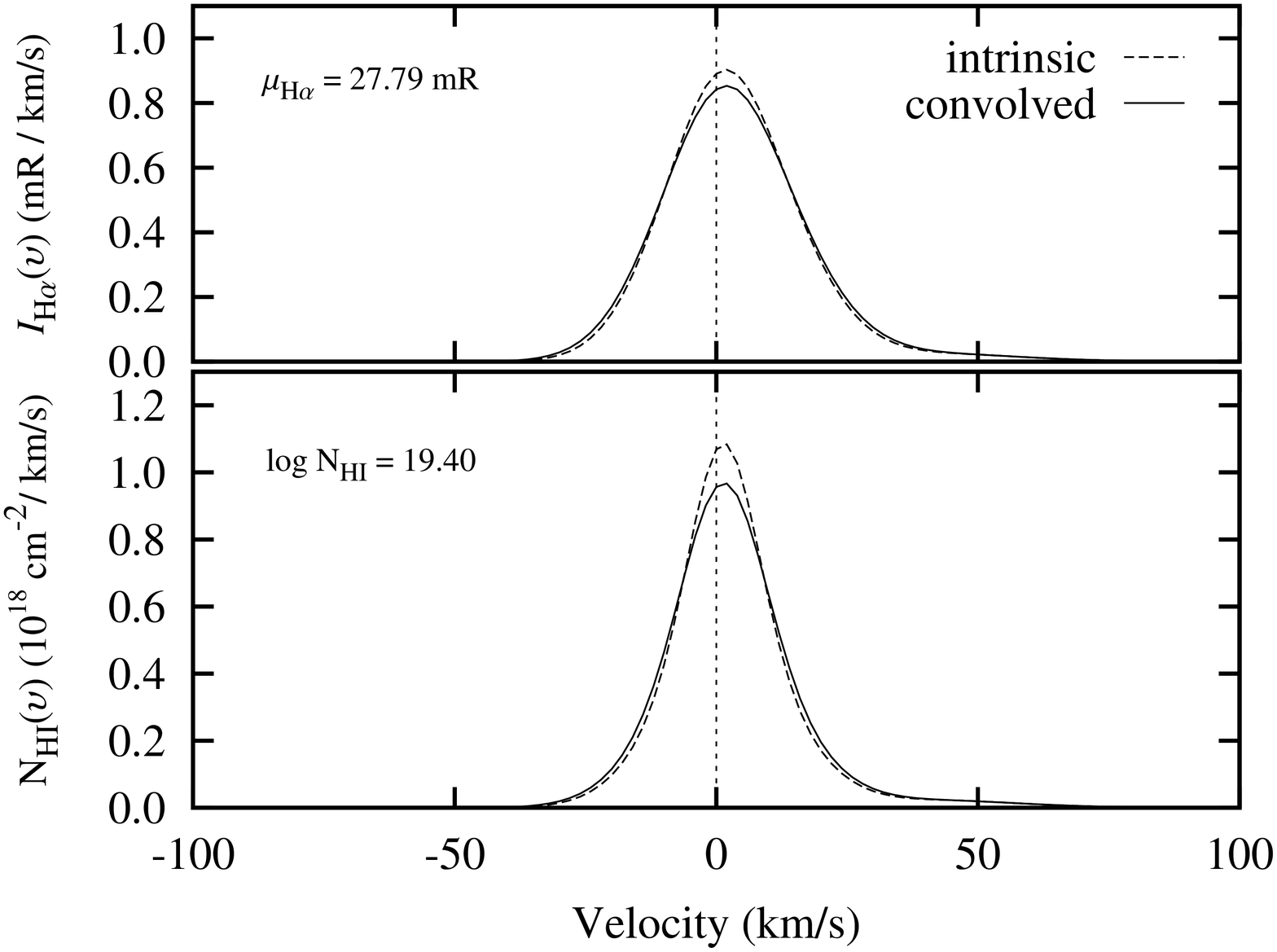}
\includegraphics[width = 0.48\textwidth]{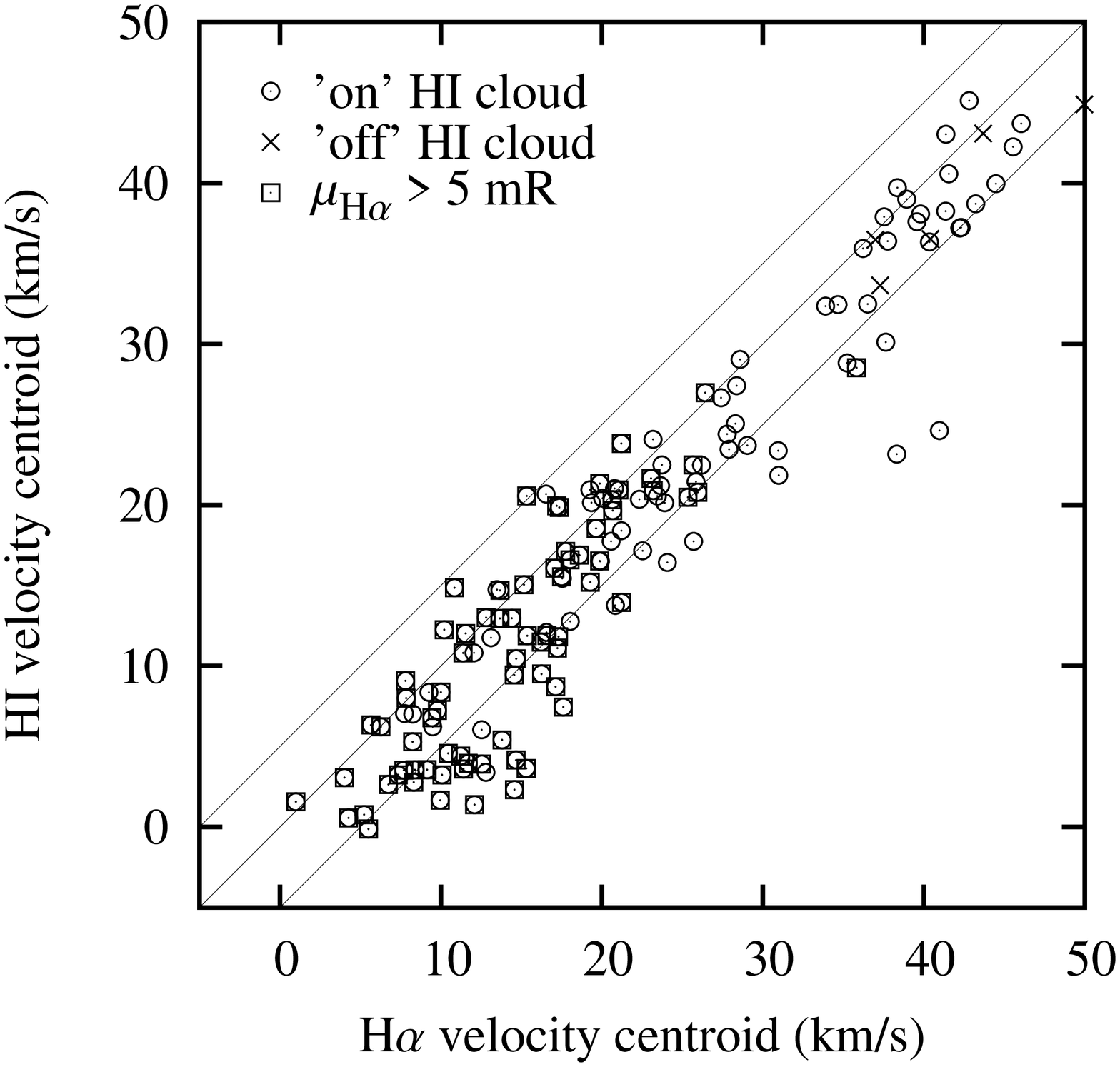}
\caption[]{Gas kinematics in model \tratio\ = 1 and \rms\ = 55 \kpc\ at $t_{sim} = 170 \Myr$. Left: Emission spectra observed at the pointing with the brightest \Ha\ emission shown in the bottom left panel of Figure \ref{fig:hamap}. We show for comparison the intrinsic line profile and the line profile convolved with a Gaussian LSF with a \mbox{FWHM = 12$\kms$}. Right: Line-of-sight velocity centroid of the \Ha\ intensity detected `on' and 'off' \HI\ clouds (see text for details). The solid diagonal lines indicate the identity line and a $\pm 5 \kms$ range around this line.}
\label{fig:kin}
\end{figure*}

\subsection{Gas kinematics}

The kinematics of the warm neutral and ionised gas phases as traced by \Ha\ and \HI\ emission provide a insight into the mechanism ionising the Stream. We explore this using the spectra of our model  identified by \tratio\ = 1 and \rms\ = 55 \kpc\ at $t_{sim} = 170 \Myr$. Note that the results, with exception of the \Ha\ emission strength, are virtually identical for all other models. An example of a strong \Ha\  and \HI\ 21cm emission lines is shown in the top panel of Figure \ref{fig:kin}. These correspond to the pointing with the brightest \Ha\ emission shown in the bottom panel of Figure \ref{fig:hamap}. The integrated strength of the line is indicated in each case in the top-left corner. For reference, the typical sensitivity of WHAM ($30 \mR$) translates into a spectral sensitivity of $0.5 \mR / \kms$, assuming a typical line width of 30 \kms. Similarly, the sensitivity of the GASS survey ($\NHI = 1.6 \times 10^{18} \psc$) corresponds to a spectral sensitivity of $10^{17} \psc / \kms$. Thus, while the \Ha\ emission in this case is marginally above the WHAM detection threshold, the corresponding \HI\ 21cm signal would be comfortably detected in a survey similar to GASS. We find typical line widths of $20 - 30 \kms$ (FWHM), which are consistent with observations \citep{put03b}.

We study the difference in the kinematics of the warm neutral and ionised gas by comparing the \los\ velocity centroid of the \Ha\ emission to the \HI\ 21cm emission, distinguishing between pointings `on' and `off' the \HI\ clouds. In this context, `on' (`off') means that the \HI\ 21cm emission is above (below) the GASS detection limit $\NHI = 1.6 \times 10^{18} \psc$. Note that the line centroid corresponds to the intensity-weighted mean velocity. In addition, we flag those pointings where the \Ha\ emission is above the level expected from ionisation by the UBV (Figure \ref{fig:kin}, bottom panel). We do not find a significant difference between the \los\ velocities of the warm neutral and ionised gas phases; their respective velocity centroids agree within $\pm 5 \kms$, as observed \citep{put03b,bar15a}. There is a tendency for the \Ha\ lines to have higher slightly higher velocity centroids. This arises from the higher line asymmetry resulting from a more extended emission along the \los. It is interesting that we barely find any \Ha\ emission above 5 mR {\em detached} from \HI\ 21cm emission. Also, the velocity of this strong \Ha\ emission is generally low ($v \lesssim 20 \kms$), with a tendency for the strongest emission to have the lowest velocities (not shown), indicating that the strong \Ha\ emission is physically associated to the \HI\ gas. This coincidence in both velocity and physical space of the \Ha / \HI\ signal is -- recall the weak dependence of the \Ha\ emission on the halo temperature -- another characteristic signature of the shock cascade.

\begin{figure}
\centering
\includegraphics[width = 0.48\textwidth]{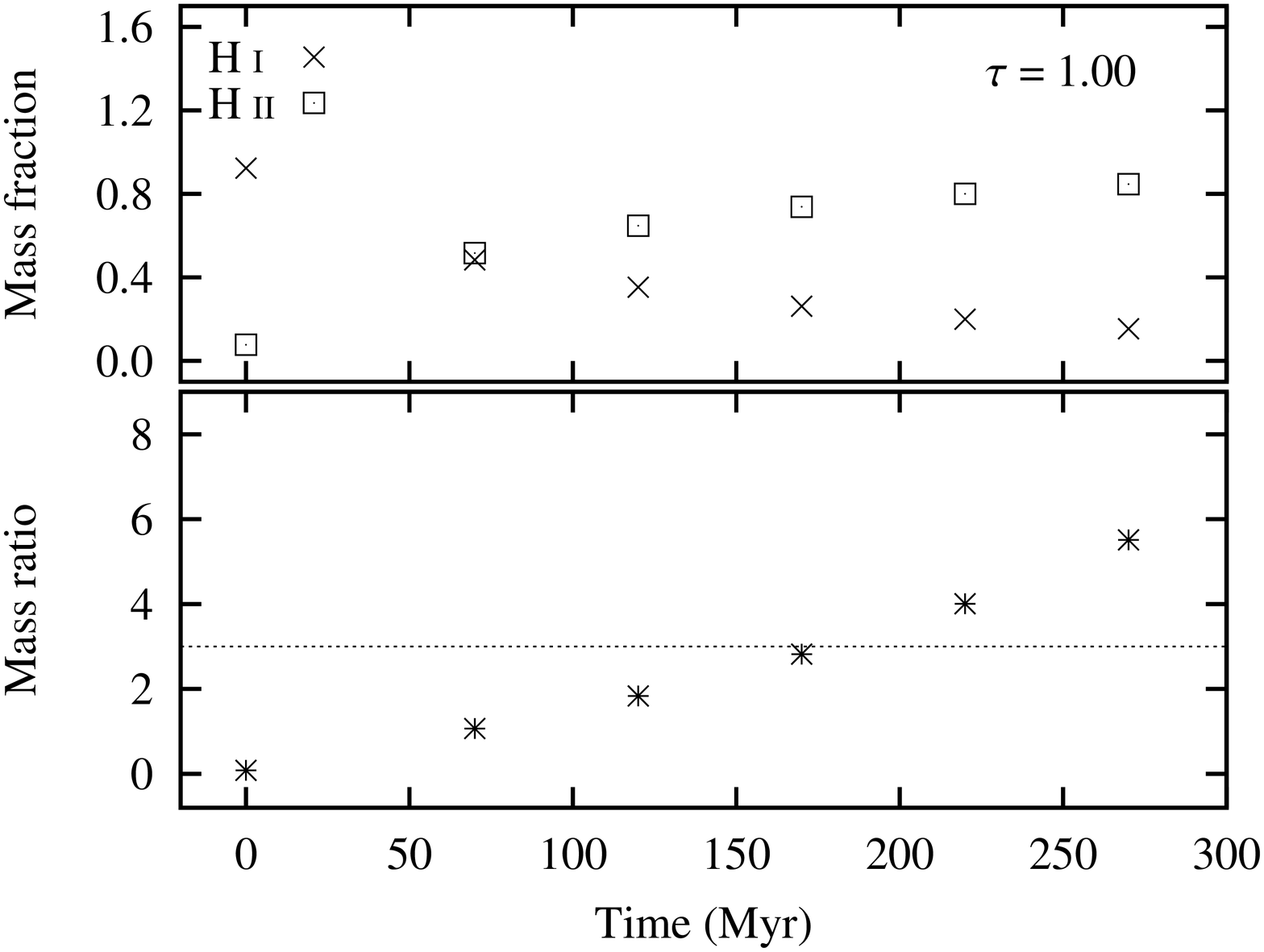}
\caption[ Evolution of the \HI\ and \HII\ mass fractions]{Evolution of the \HI\ and \HII\ mass fractions in the \tratio\ = 1 model Stream at \rms\ = 55 kpc,  projected along the $y$-axis. The top panel shows the evolution of the individual mass fractions $M_X / M_{tot}$, where $X \in \{ \HI, \HII \}$ and $M_{tot} = M_{\!\HI|} + M_{\!\HII}$. The bottom panel shows the evolution of the \HII\ to \HI\ mass ratio. The dashed horizontal line indicates the ratio \HII\ : \HI\ = 3 inferred from observations \citep{fox14a}. The results are essentially the same for all other models, and are therefore not shown.}
\label{fig:ionfracevol}
\end{figure}

\subsection{Gas ionisation timescales} \label{sec:ion}

It has recently been inferred that the mass of the ionised gas kinematically associated to the Magellanic Stream is roughly 3 times larger than its \HI\ mass \citep{fox14a}. Here, we briefly explore the evolution of the ionised and warm-neutral gas mass fractions, using the \HII\ and \HI\ masses in as a proxy for the ionised and warm-neutral gas phases, respectively. Again, we focus on the results obtained from our models characterised by \tratio\ = 0.50, 0.75, 1.00 and \rms\ = 55 \kpc. Note that the results are qualitatively the same for all other models. Figure \ref{fig:ionfracevol} shows the evolution of the individual mass fractions $f_X \equiv M_X / M_{tot}$, where $X \in \{ \HI, \HII \}$ and $M_{tot} = M_{\!\HI} + M_{\!\HII}$, as well as the evolution of the \HII\ to \HI\ mass ratio. We include for reference the inferred value of the ratio \HII\ : \HI\ $= 3$. We find that the gas evolves on a typical timescale of 100 \Myr, which is consistent with the time estimate which results from assuming that the cloud disrupt by the action of Kelvin-Helmholtz instabilities and our adopted (initial) value of $\eta$. After this period, the ionisation effect of the halo-cloud and cloud-cloud interaction leads to a reduction of the warm neutral gas mass by half. The \HII\ : \HI\ mass ratio increases rapidly with time, and the inferred \HII\ : \HI\ mass ratio of 3 is reached after $\sim170 \Myr$.

Since our simulations assume free boundary conditions, it is difficult to quantify with precision how much of the neutral gas is ionised and how much simply escapes the simulation volume. Nonetheless, we estimate that only a negligible fraction of the warm-neutral gas is lost by $t_{sim} \lesssim 270 \Myr$. Therefore, the ionisation of the gas due to interactions with the hot halo gas, and due to cloud-cloud interactions lead to a strong evolution of the mass fractions in the neutral and ionised phases. The relatively short survival timescale implies the requirement for a continuous replenishment of gas from the MC to the Stream \citep{bla07a}.\\
 
Note that, for the ease of discussion, in the following we shall refer to the set of models discussed in the last sections, defined by $n(55 \kpc) = 2 \times 10^{-4} \pcc$, as the `standard' model set.

\subsection{Conservative departures from the standard models} \label{sec:modvars}

\begin{figure*}
\centering
\includegraphics[width = 0.48\textwidth]{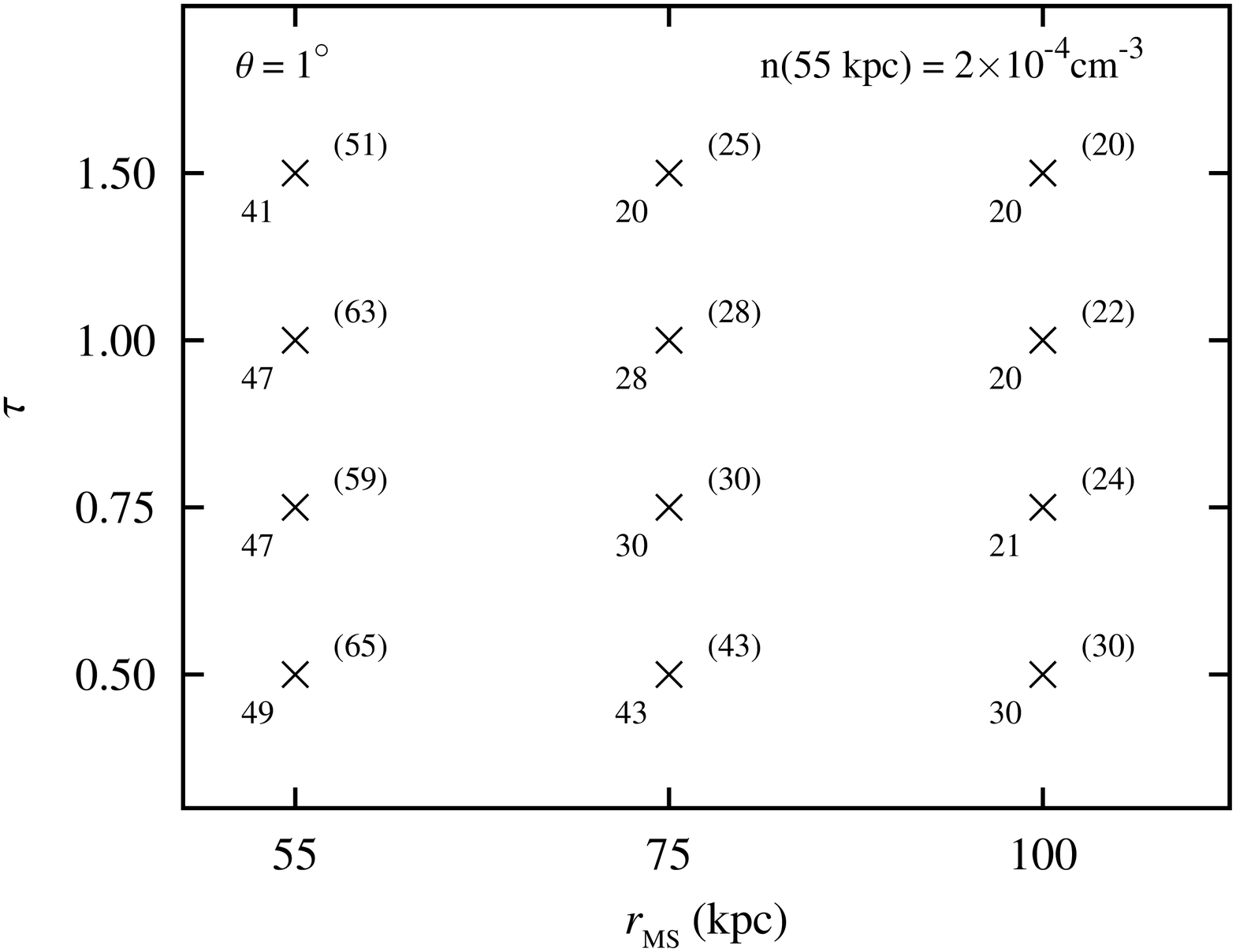}
\includegraphics[width = 0.48\textwidth]{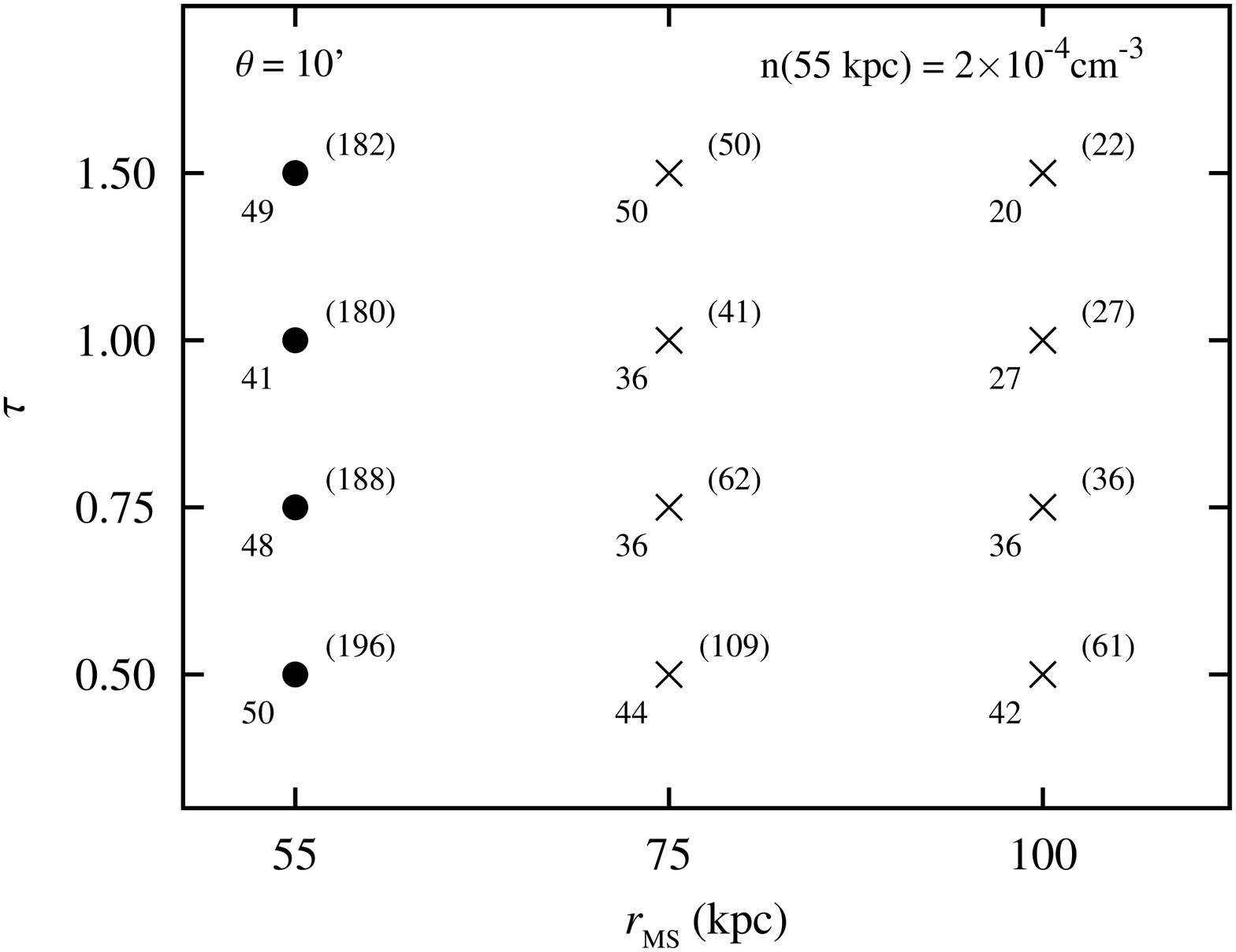}\\
\includegraphics[width = 0.48\textwidth]{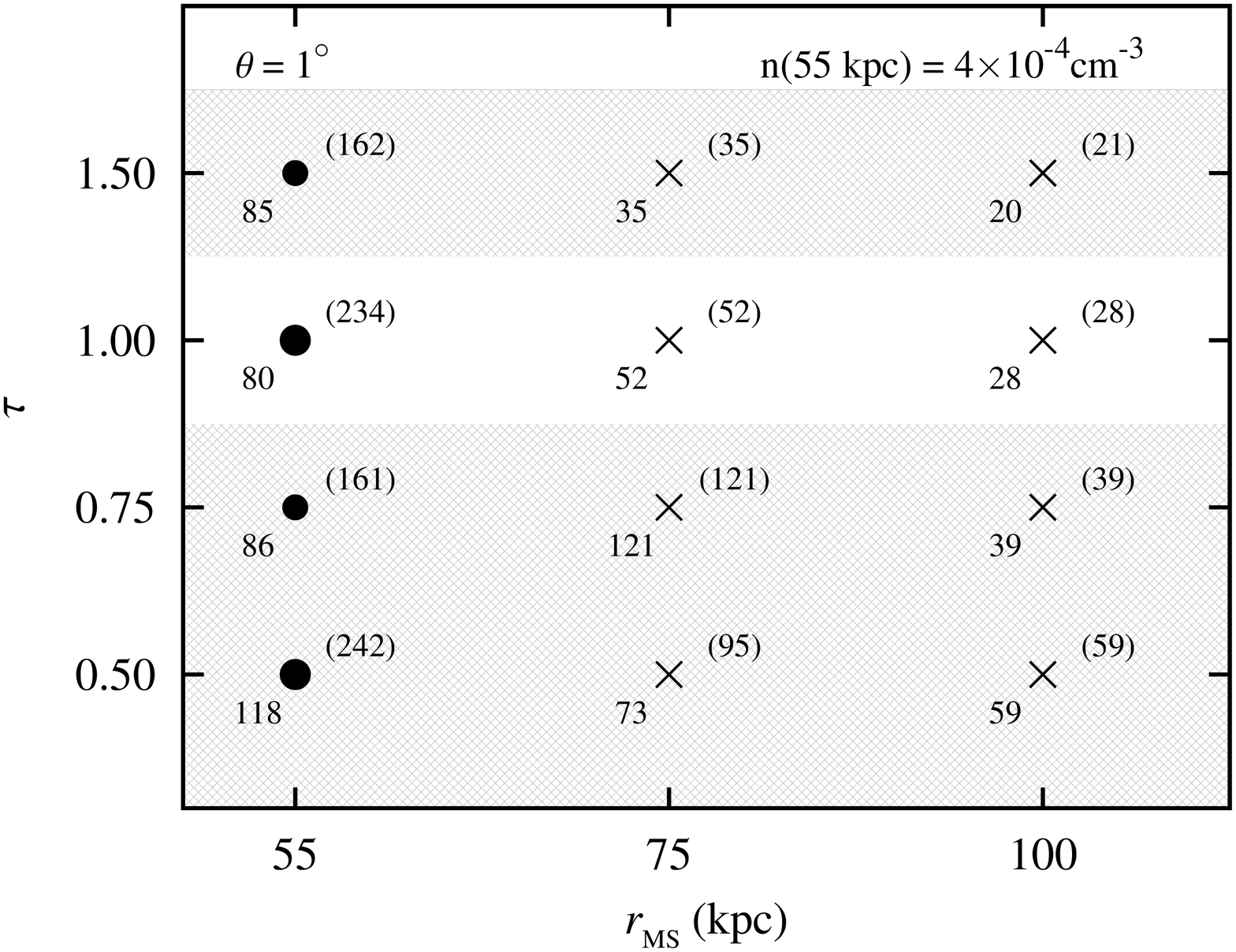}
\includegraphics[width = 0.48\textwidth]{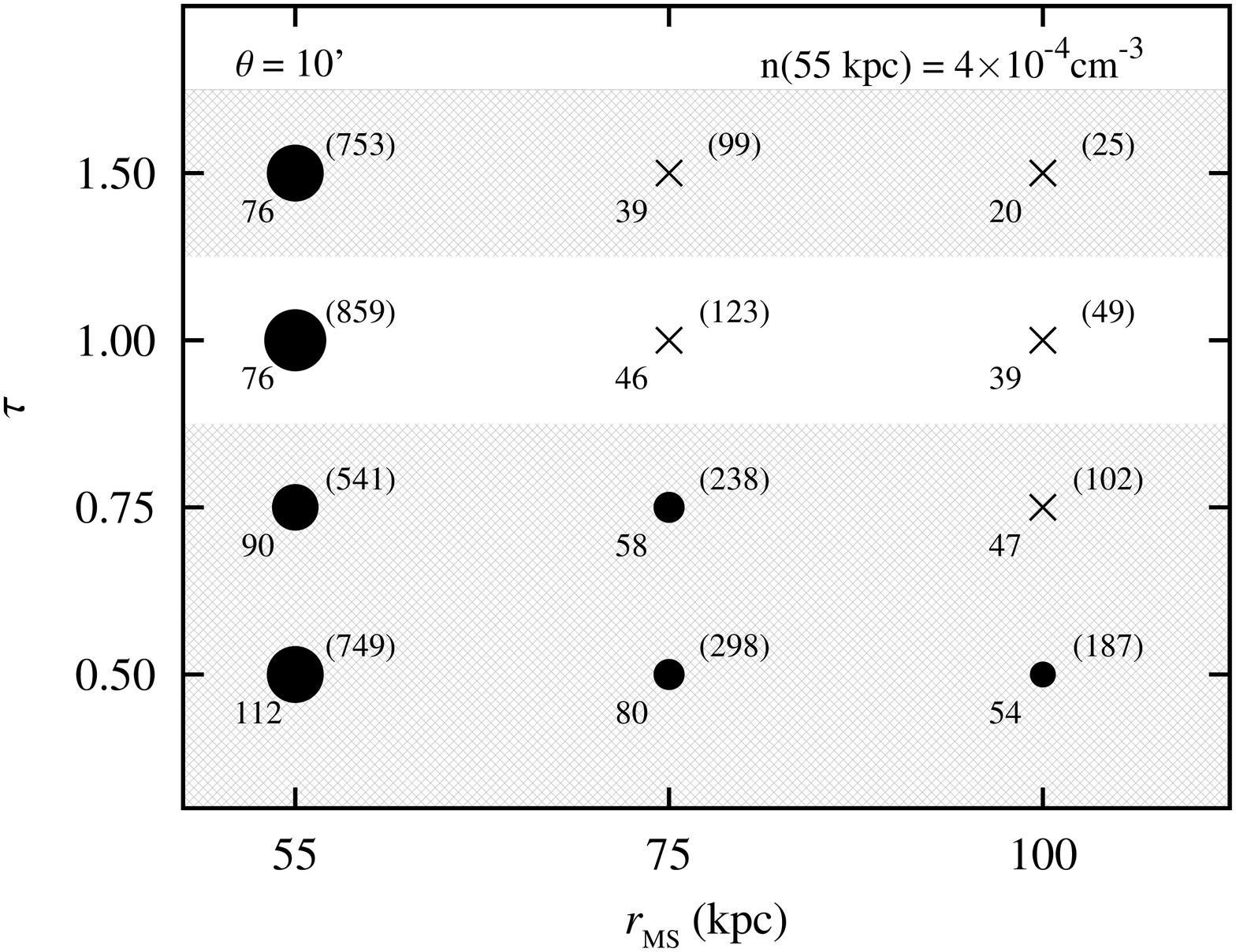}
\caption[]{Overview of model results. In any panel, a model is identified by the set of parameters \{$n(55 \kpc)$, $\theta$, \tratio, \rms\}. The corresponding data point consists of a symbol (\textbullet\ or $\times$), and two numerical values. The value to the top-right of a given symbol (in parentheses) indicates the {\em maximum} \Ha\ emission (mR) observed in the full time lapse of 320 Myr in that particular model, while the value to the bottom-left indicates the \Ha\ emission (mR) at the 90 percentile level. A circle (\textbullet) indicates whether the maximum \Ha\ intensity exceeds 160 mR; a cross ($\times$) signals failure to do so. The circle diameter is roughly proportional to the maximum \Ha\ intensity in each case. The grey hatched area highlights models that are inconsistent with the halo mass constraint (see text for details). Note that the top (bottom) panels correspond to models where $n(55 \kpc) = 2 \times 10^{-3} \pcc$ ($4 \times 10^{-3} \pcc$), and left (right) panels to models where a low (high) resolution beam with diameter $\theta = \onedeg$ (10') has been adopted to produce virtual observations. All models at $\rms > 100 \kpc$ have been omitted since their respective \Ha\ emission is weak.}
\label{fig:modvars}
\end{figure*}

Given the results of the last sections, our standard set of models appear to indicate that the shock cascade {\it fails} to produce the mean level of \Ha\ emission \citep[$\sim160 \mR$;][]{bla13a} observed along the Magellanic Stream. But there are two factors
that deserve closer consideration. First, a halo density at any given distance within \rvir\ {\em higher} than implied by our standard models could enhance the onset and development of hydrodynamic instabilities (Kelvin-Helmholtz), thus promoting the shock cascade and the resulting \Ha\ emission. Secondly, an increase in the beam resolution would certainly diminish the smearing effect on bright spots which have characteristic sizes significantly smaller than the beam. Consider, for instance, that a beam with a diameter of 10' samples a region which is nearly ten times smaller than a \onedeg\ beam at \rms\ = 55 kpc. Indeed, the brightest \Ha\ observations along the Stream have been obtained with spectroscopy over smaller apertures (3'-10') than the WHAM survey \citep[e.g.][]{put03b}.

Therefore, it is important to extend the parameter space of our study, in terms of both the halo density and the adopted beam size. We now {\em increase} the normalisation of the halo gas density at 55 kpc by a factor of 2, i.e. $n(55 \kpc) = 4 \times 10^{-4} \pcc$, but with initial conditions and set up which are otherwise identical in every aspect to the standard models. These shall be referred to as the 'extended' models. In addition to increasing the density, we produce a new set of virtual observations for the standard models, adopting a smaller beam size (i.e. a higher resolution) with a diameter of $\theta = 10'$ (rather than \onedeg), and two sets of virtual observations for the extended models, adopting either a low (\onedeg) or a high (10') resolution beam. We shall refer to these as the `low-resolution' and `high-resolution' models, respectively, keeping in mind that is not the actual hydrodynamical model, but the virtual observation, to which the resolution refers.

Note that the density profiles implied by the extended models are fairly consistent with observations (Figure \ref{fig:dens}, bottom-left panel), although there is no model which agrees with the data over the full range in distance. However, the enclosed gas mass in the halo within \rvir\ resulting from each of these density profiles is larger than the the mean range of masses inferred from observations, and - with exception of the \tratio\ = 1 extended model -- they are all inconsistent with the mass limit imposed by the universal baryon-to-total-mass fraction (Figure \ref{fig:dens}, bottom-right panel). Hence, all the extended models but  the \tratio\ = 1 model, may be deemed 'unphysical'. Nonetheless, it is still of interest to explore the \Ha\ intensity in these type of models, as will be discussed later.

Each of the models in either the standard or the extended set is run for a total simulation time of $t_{sim} = 320 \Myr$, and a virtual observation of the \Ha\ intensity of the gas at the appropriate resolution is produced every $\Delta t_{sim} = 10 \Myr$. Therefore, for each model and at each time step we obtain a whole new distribution of \Ha\ intensities. In order to deal with the overwhelming amount of information, and to make a meaningful comparison between models, we opt for the following approach: Since the total time lapse $t_{sim}$, and the choice of output time step are somewhat arbitrary, for each model we single out the snapshot at which the maximum \Ha\ intensity anywhere in the gas (i.e. at any beam pointing) is largest. This is further justified by the fact that the maximum \Ha\ emission does not evolve strongly with time (Figure \ref{fig:haevol}). Note that the snapshot thus selected will in general be different for each model. For this particular snapshot and model, we also obtain the value of the emission at the 90 percentile level of the corresponding distribution. We then assess the performance of each model simply by comparing both the maximum \Ha\ emission to the mean level of \Ha\ emission ($\sim 160 \mR$) observed along the Magellanic Stream. The result of these approach applied to both the standard and the extended models is collected in Figure \ref{fig:modvars}.

There, the top panels correspond to the standard models, and the bottom panels show the results for the extended set; the left (right) panels correspond to the low (high) resolution cases. In any panel, each `data point' corresponds to a particular model identified by \{$n(55 \kpc)$, $\theta$, \tratio, \rms\}, and it consists of a symbol (circle or cross), and two numerical values. The value to the top-right of a given symbol (in parentheses) indicates the {\em maximum} \Ha\ emission (in mR), while the value to the bottom-left indicates the value at the 90 percentile level. A circle indicates whether the maximum \Ha\ intensity exceeds 160 mR; a cross signals failure to do so. The circle diameter is roughly proportional to the maximum \Ha\ intensity in each case. In addition, we have greyed hatched the parameter space corresponding to models that are deemed 'unphysical' as per the above discussion. As a guide, note that the series of numbers in parentheses, i.e. the maximum \Ha\ intensity, shown on the top-left panel for model \tratio\ = 1 correspond to the results shown in the top-right panel of Figure \ref{fig:haevol} at 220 Myr (ignoring the 150 kpc series). Similarly, the maximum \Ha\ intensities at \rms = 55 kpc for all models correspond to the results shown in the bottom-right panel of Figure \ref{fig:haevol} at 170 Myr \tratio\ = 0.5, and at 220 Myr for all the other models.

Apparently, both a higher density and an improved resolution enhance the \Ha\ emission, but in different ways. On the one hand, increasing the density shifts the overall \Ha\ intensity towards the high-end. This can be seen by comparing the maximum \Ha\ intensity (and the 90 percentile) between the standard and the extended models, which are roughly a factor 2 - 3 higher in the latter. Increasing the resolution, on the other hand, boosts only the maximum \Ha\ emission, without significantly affecting the distribution of intensities as a whole, as can be judged by comparing the values at the 90 percentile level.

The standard models at low resolution (top-left panel) fail dramatically at any distance in matching the Stream's mean emission. Both the standard model at high resolution (top-right) and the extended model at low resolution (bottom-left) result in \Ha\ intensities in the 100 mR regime only in the near field at $55 \kpc$. In this sense, both a higher halo density model, and virtual observations with a high resolution are equally crucial factors in pushing the \Ha\ emission towards higher levels. It is worth mentioning at this point that the original model by \cite{bla07a} - an instance of a `standard' model in our terminology -- implicitly assumed an infinite resolution, and was therefore capable of reproducing \Ha\ emission at levels of a few hundred milli-Rayleigh at 55 kpc.

The extended, high resolution models (bottom-right) are the most promising of all the models considered here. In the near field, the shock cascade in all these models results in \Ha\ intensities which reach, or even exceed, the highest levels of $\sim 700 \mR$ observed along the Magellanic Stream over the SGP. In the far field ($\!\rms\ \sim 75 - 100 \kpc$), two of the models (\tratio\ = 0.5 and 0.75) produce \Ha\ emission consistent with the mean emission of $\sim 160 \mR$ observed along the Stream. However, the success of these models comes at a cost. The increase in the halo density is accompanied by an increase in the halo mass. This makes all but the \tratio\ = 1 model be {\em inconsistent} with the limit on the gas mass of the Galaxy imposed by the cosmic fraction of baryons relative to the total mass. Given that the isothermal sphere model yields an upper limit on the halo gas density at any distance and for a given gas mass, the situation is even more unfortunate for any other reasonable DM halo model.

\section{Discussion}

Within the context of the `shock cascade' model, we have shown that the interpretation of the MS optical emission ($\approx 100 - 200$ mR) away from the Magellanic Clouds may still work for the updated parameters of the Galaxy under a narrow set of conditions. Conventionally, the MS was assumed to be on a circular orbit at the midpoint of the LMC and SMC ($d\approx 55$ kpc). At this distance, for a smooth halo density profile, as long as the coronal halo density satisfies $2 \times 10^{-4} < (n / \pcc) < 4 \times 10^{-4}$, the shock cascade generates sufficient \Ha\ emission to explain the observations. The upper limit on density ensures that the mass of the corona (when including the other baryonic components) does not exceed the baryonic mass budget of the Galaxy.

In recent years, the first accurate measurements of the MC's proper motions, combined with a smaller estimate
of the Galaxy's total mass, have led to a major revision of their binary orbit about the Galaxy. A highly elliptic orbit 
is now favoured by most researchers, which pushes the Stream's mean distance further out than the conventional
assumption. For a smooth halo, the lower density limit of the above range can occur at $d \approx 75$ kpc without 
violating the constraint on the baryonic mass budget. If the Stream's mean distance (especially over the SGP) happens to exceed this limit,
then we are forced to either reject the model, or consider more complex density distributions for the Galactic corona (see below).

An alternative interpretation of the Stream's optical emission has recently been put forward in the context of the Galaxy's nuclear activity. The energetic bubbles observed with the Fermi-LAT in gamma rays \citep{su10a} indicate that a powerful event has taken place at the nucleus in the recent past.  If the gamma rays are produced through inverse Compton upscattering of soft photons, this event can be dated to 1-3 Myr ago \citep{guo12a}. Within the context of this model, \citet{bla13a} show that the Stream \Ha\ emission can also be explained by accretion-disk driven ionization for Stream distances of 100 kpc or more over the poles. The recent discovery of high ionization species (e.g. \SiIV, \CIV) over the SGP \citep{fox15a} may lend further support to this model. On the other hand, \citet{bla07a} provide diagnostics of slow shocks (e.g. Balmer decrement) that are likely to be observable along the Magellanic Stream in future observing campaigns. If enhanced Balmer decrements ($\!\Ha/ \!\Hb \gtrsim 3$) are confirmed along the Stream, then some variant of the shock cascade model may be needed.

There is now increasing evidence that the CGM of low redshift galaxies is multi-phase, with a comparable fraction of baryons both in a hot and a warm phase \citep{wer14a}. Modern simulations of the CGM also suggest that the hot halos of galaxies are likely to be heavily structured, at least during a major phase of gas accretion. While CDM accretion may be isotropic on average, individual events involving massive systems are not, as clearly demonstrated by the Magellanic Clouds in orbit around the Galaxy, or the Sgr dwarf which extends through much of the halo \citep{iba94a}. The mass of this system was probably comparable to the LMC and may well have retained gas before being tidally disrupted. A more recent accretion event is attested by the massive \HI\ stream, the Smith Cloud ($\gtrsim 2\times 10^6$\Msun), that is presently being stripped and ablated by the corona \citep{bla98a,loc08a,nic14b}.

We may need therefore to consider the possibility that the corona is inhomogeneous rather than smooth. This would allow for significant density variations along different directions and at different distances, without violating constraints on the total baryonic mass of the Galaxy. In turn, based on our extended models, this would restore the shock cascade as a viable model to explain the \Ha\ emission, allowing the Stream to lie at $\sim 100$ kpc the SGP as predicted by dynamical models \citep[e.g.][]{gug14a}. In this scenario, we may envisage the strong optical emission along the Stream as a result of the Stream's gas colliding with high-density debris of past accretion events scattered along its orbit. At this point, however, it is not clear what a suitable model for an inhomogeneous hot halo might be. We will address this in future work.

If confirmed, the larger distance to the Stream of 100 kpc would lift its mass to roughly $8 \times 10^9 \Msun$, comparable to the total coronal gas mass. If it all breaks down, it would roughly double the mass of the corona, at least for a while. We thus speculate that the halo of the Galaxy is substantially mass-loaded with gas lost by smaller accreted systems. The interaction with the hot halo may prevent this gas from cooling sufficiently to condense and `rain' down on the disc. Such a process is analogous to the meteorologic phenomenon know as {\em virga}, a type of atmospheric precipitation that evaporates while dropping and thus fails to reach the ground. The heavy halo may thus serve as a huge reservoir, from which gas may eventually be forced out by the strong interaction at the disc-halo interface \citep[e.g.][]{mar10b}.

\acknowledgments

We thank the anonymous referee who made a number of excellent suggestions which significantly improved the presentation of our results. TTG acknowledges financial support from the Australian Research Council (ARC) through a Super Science Fellowship and an Australian Laureate Fellowship awarded to JBH.

\bibliographystyle{apj}
\bibliography{/Users/tepper/references/complete}

\newpage

\appendix

\section{\Ha\ recombination coefficient} \label{sec:harec}

We describe the temperature dependence of the hydrogen {\em total} case B recombination coefficient, $\alpha_{B}$, and of the effective \Ha\ recombination coefficient, $\alphabha $, with the generic fitting formula \citep[][]{peq91a}:
\begin{equation} \label{eq:peqeff}
	\alpha (T) = \alpha_0 \times \, \frac{ (1 + c) (T / 10^4 K)^b}{1 + c ~(T / 10^4 K)^d} \, .
\end{equation}
The parameter values appropriate in each case are, respectively, $\alpha_0 = 2.585  \times 10^{-13} \cm^3 \ps$, $b = -0.6166$, $c = 0.6703$, and $d = 0.5300$, and $\alpha_0 = 1.169  \times 10^{-13} \cm^3 \ps$, $b = -0.648$, $c = 1.315$, and $d = 0.523$. Note that this formula is accurate to two percent in the range $40 \K < T < 2 \times 10^4 \K$.

\section{Photoionisation-induced \Ha\ emission} \label{sec:pi}

We include in our model the contribution of the cosmic UV background (UVB) ionising radiation, which leads to a low, but non-negligible level of \Ha\ emission along the Stream. For simplicity, we assume that the gas is highly ionised ($n_{\rm H} \approx n_{\!\HII}$); and photoionisation {\em equilibrium}, which implies that the ionisation and recombination events balance each other:
\begin{equation} \label{eq:pieq}
	(n_e ~n_{\!\HII} ) \; \alpha_B = n_{\!\HI} \; \Gamma_{\!\HI} \, ,
\end{equation}
Here, $\Gamma_{\!\HI}$ is the \HI\ photoionisation rate in units of photon per atom per second; $J_{\nu}$ is the angle-averaged specific intensity of the UVB; $\nu_{ \textnormal{\sc ll} } \approx 3.29 \times 10^{15} \Hz$ is the minimum photon frequency required to ionise hydrogen; and $\sigma_{\rm H}(\nu)\approx \sigma_0 \; (\nu / \nu_{ \textnormal{\sc ll} })^3$ is the hydrogen ionisation cross section with $\sigma_0 = 6.3 \times 10^{-18} \cm^{2}$.

The \Ha\ emission induced by the metagalactic ionising radiation field along the \los\ is thus given by equation \eqref{eq:haphot}, where the fraction of recombinations that produce an \Ha\ photon is
\begin{equation} \label{eq:fha}
	f_{\!\Ha}(T) \equiv \frac{ \alpha_{B}^{ (\textnormal{\Ha}) } (T)}{ \alpha_B(T) } \approx 0.452 ~g(T)\, ,
\end{equation}
Here, $g$ is a monotonically decreasing function of temperature (see equation \ref{eq:peqeff})
\begin{equation} \label{eq:gt}
	g(T) = \left( \frac{ 1 + c }{ 1 + c' } \right) \left[ \frac{ 1 + c' ~(T / 10^4 K)^{d'} } { 1 + c ~(T / 10^4 K)^{d} } \right] (T / 10^4 K)^{b - b'} \, ,
\end{equation}
which satisfies $g(10^4 \K) \equiv 1$, and $g \in (0.6, \, 1.3)$ for $T \in [10^3, \, 10^6] \K$.

In general, $f_{\!\Ha}$ (through $T$) and $n_{\!\HI}$ both vary along the \los. However, an estimate of the \Ha\ signal resulting from the ionisation by the cosmic UVB of gas at can nevertheless be obtained assuming the gas temperature to be uniform along the \los. In this case, and inserting the appropriate numerical values we get
\begin{equation}
	\mua^{(phot)} \approx 452 \mR ~\left( \frac{ \NHI }{ 10^{18} \psc} \right) ~\left( \frac{ \Gamma_{\!\HI} }{ 10^{-12} \ps} \right) \, ,
\end{equation}
where $\NHI$ is the integral of $n_{\!\HI}$ along the \los. Hence, for an \HI\ ionisation rate $\Gamma_{\!\HI} \sim 10^{-13} \ps$ at $z = 0$ and gas at $10^4 \K$ with $\NHI \sim 10^{17} \psc$, the \Ha\ signal resulting from the ionisation by the cosmic UVB of gas at $10^4 \K$ is roughly 5 mR.

Since we are not performing proper radiative transfer calculations, we limit the depth (along the \los) of the gas ionised by the UVB to a value $L_{max}$ defined by the condition that the column recombination equals the incident ionising photon flux $\psi_i$:
\begin{equation} \label{eq:pieq2}
	\alpha_B  \int_{0}^{L_{max}} \!\!\!\!\!\!\!\! (n_e ~n_{\!\HII} ) ~ds \stackrel{!}{=} \psi_i \, ,
\end{equation}
where the ionising photon flux  (in photons \psc \ps) is given by equation \eqref{eq:ionflux}. This condition implies that all the ionising photons be absorbed within a depth $L_{max}$, assuming the gas has been exposed to the (uniform) UV radiation field long enough to reach ionisation equilibrium \citep[which is well justified in the case of the Stream; see][their Appendix]{bla13a}. Note that this condition is equivalent to restricting the ionising effect of the UVB to a column of gas $\sim 10^{17} \psc$. Indeed, using equations \eqref{eq:pieq} and \eqref{eq:ionflux}, the condition \eqref{eq:pieq2} becomes (using $\gamma = 1.8$)
\begin{equation}
	\int_{0}^{L_{max}} \!\!\!\!\!\!\!\! n_{\!\HI} ~ds = \frac{ 2 }{ 3 } \NHI({\rm LL}) \, . \notag
\end{equation}

where $\NHI({\rm LL}) \equiv \sigma_0^{-1} \approx 1.6 \times 10^{17} \psc$.\\


\end{document}